\documentclass[aps,prx,twocolumn,superscriptaddress]{revtex4-1}

\usepackage{graphicx}
\usepackage{amsmath}
\usepackage{multirow}
\usepackage{array}
\newcommand{\sio}{Sr$_2$IrO$_4$}
\newcommand{\ppp}{($\pi$-$\pi'$) }
\newcommand{\psp}{($\pi$-$\sigma'$) }
\newcommand{\spp}{($\sigma$-$\pi'$) }

\begin{document}

\title{Pseudospin-lattice coupling in the spin-orbit Mott insulator \sio
}

\author{J. Porras}
\author{J. Bertinshaw}
\author{H. Liu}
\author{G. Khaliullin}
\author{N. H. Sung}
\affiliation{Max-Planck-Institut f{\"u}r Festk{\"o}rperforschung, Heisenbergstra{\ss}e 1, D-70569 Stuttgart, Germany}
\author{J.-W. Kim}
\affiliation{Advanced Photon Source, Argonne National Laboratory, Argonne, Illinois 60439, USA}
\author{S. Francoual}
\affiliation{Deutsches Elektronen-Synchrotron DESY, 22603 Hamburg, Germany}
\author{P. Steffens}
\affiliation{Institut Laue-Langevin 6, rue Jules Horowitz, BP 156, 38042 Grenoble Cedex 9, France}
\author{G. Deng}
\affiliation{Australian Nuclear Science and Technology Organization, Lucas Height, NSW 2233, Australia}
\author{M. Moretti Sala}
\affiliation{European Synchrotron Radiation Facility, BP 220, F-38043 Grenoble Cedex, France}
\affiliation{Dipartimento  di  Fisica,  Politecnico  di  Milano,  Piazza  Leonardo  da  Vinci  32,  I-20133 Milano, Italy}
\author{A. Efimenko}
\affiliation{European Synchrotron Radiation Facility, BP 220, F-38043 Grenoble Cedex, France}
\author{A. Said}
\author{D. Casa}
\author{X. Huang}
\author{T. Gog}
\author{J. Kim}
\affiliation{Advanced Photon Source, Argonne National Laboratory, Argonne, Illinois 60439, USA}
\author{B. Keimer}
\affiliation{Max-Planck-Institut f{\"u}r Festk{\"o}rperforschung, Heisenbergstra{\ss}e 1, D-70569 Stuttgart, Germany}
\author{B. J. Kim}
\email{bjkim6@postech.ac.kr}
\affiliation{Max-Planck-Institut f{\"u}r Festk{\"o}rperforschung, Heisenbergstra{\ss}e 1, D-70569 Stuttgart, Germany}
\affiliation{Department of Physics, Pohang University of Science and Technology, Pohang 790-784, South Korea}
\affiliation{Center for Artificial Low Dimensional Electronic Systems, Institute for Basic Science (IBS), 77 Cheongam-Ro, Pohang 790-784, South Korea}
\date{\today}

\begin{abstract}
Spin-orbit entangled magnetic dipoles, often referred to as pseudospins, provide a new avenue to explore novel magnetism inconceivable in the weak spin-orbit coupling limit, but the nature of their low-energy interactions remains to be understood. We present a comprehensive study of the static magnetism and low-energy pseudospin dynamics in the archetypal spin-orbit Mott insulator \sio. We find that in order to understand even basic magnetization measurements, a formerly overlooked in-plane anisotropy is fundamental. In addition to magnetometry, we use neutron diffraction, inelastic neutron scattering and resonant elastic and inelastic x-ray scattering to identify and quantify the interactions that determine the global symmetry of the system and govern the linear responses of pseudospins to external magnetic fields and their low-energy dynamics. We find that a pseudospin-only Hamiltonian is insufficient for an accurate description of the magnetism in \sio\, and that pseudospin-lattice coupling is essential. This finding should be generally applicable to other pseudospin systems with sizable orbital moments sensitive to anisotropic crystalline environments.
\end{abstract}

\maketitle

\section{Introduction}
 4$d$ and 5$d$ transition-metal compounds are characterized by spin-orbit entangled and spatially-extended valence electrons, which in magnetic insulators translate to strong and long-ranged interactions among pseudospins. Pseudospins, having sizable orbital contributions to the magnetic moment, are highly sensitive to the crystalline symmetry~\cite{Khaliullin2005}, and thus interact through multiple interactions whose hierarchy depends on the lattice geometry and the pseudospin quantum number. For example, pseudospins-1/2 in a honeycomb lattice have dipolar-like, bond-directional interactions, which dominate over isotropic (Heisenberg) interactions and constitute the key building block for the Kitaev spin liquid~\cite{Jackeli,Chal10,Choi12,Chal13,Hwan15}. The opposite is true for a square lattice in which the leading order interaction is isotropic, rendering a rare realization outside of the cuprate family of a (pseudo)spin-1/2 antiferromagnet on a square lattice~\cite{Kim2008,Kim2009,Kim2012,Fujiyama,Bertinshaw2018}. Pseudospins-1 on the same lattice may be subject to a single-ion anisotropy that is much stronger than all nearest-neighbor (NN) interactions and leads to distinct physics characterized by ``soft'' magnetic moments supporting a Higgs amplitude mode~\cite{Khaliullin2013,Jain2017}.

On the experimental side, recent technological advances in resonant inelastic x-ray scattering (RIXS)~\cite{note1} have allowed key insights into the nature of magnetism expressed by pseudospins through measurement of the momentum-resolved dynamic structure factor. Extensive efforts in the last several years have revealed the nature of leading-order interactions in a number of strongly spin-orbit coupled materials: e.g. bond-directional interactions in Na$_2$IrO$_3$~\cite{Hwan15}, Heisenberg interactions in Sr$_2$IrO$_4$~\cite{Kim2012,GretarssonRIXS,Pincini}, and Ising interactions in Sr$_3$Ir$_2$O$_7$~\cite{PhysRevLett.109.157402}. However, the limited energy resolution of RIXS has so far not allowed for substantial information beyond the leading-order interactions. Despite their smaller energy scales, next-order interactions play a crucial role in determining the magnetic phase of the system and its stability against perturbations. For instance, the Kitaev spin liquid phase has a finite window of stability when perturbed by Heisenberg interactions~\cite{Chal10}. For magnetically ordered systems, the low-energy physics determines the global symmetry of the magnetic structure and thereby the topology of the electronic system as a whole. A prominent example is the pyrochlore iridates with the so-called all-in-all-out magnetic structure, which is a prerequisite for the Weyl semi-metal phase predicted in Nd$_2$Ir$_2$O$_7$~\cite{Wan2011}.

\begin{figure*}
\includegraphics[width=0.90\linewidth]{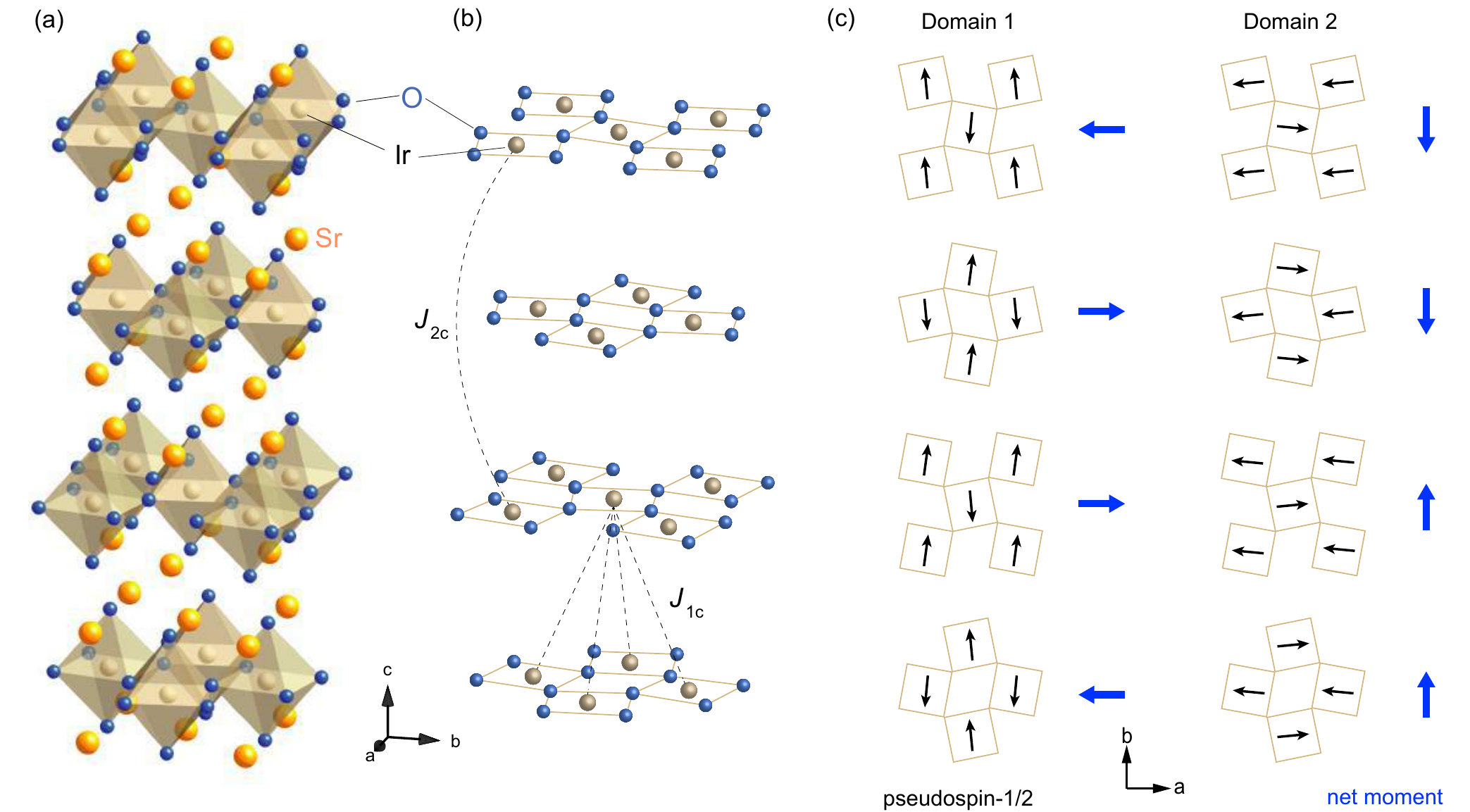}
\caption{\label{fig1}(a) The  crystal structure of \sio\  is tetragonal (space group $I4_1/a$) with $a=b=5.49\ \text{\AA}$ and $c=25.80\ \text{\AA}$ at room temperature. The tetragonal $a$ and $b$ axes are rotated by 45 degrees from the Ir-O-Ir bond directions. Ir atoms lie in the center of oxygen octahedra. IrO$_2$ layers are separated by SrO layers. (b) interlayer pseudospin couplings between the nearest layers and the next-nearest layers via $J_\mathrm{1c}$ and $J_\mathrm{2c}$, respectively. (c) Top view on the IrO2 planes, with arrows indicating canted pseudospins (black) and net ferromagnetic moments (blue), following the possible magnetic domain configurations in the 4$_1$ crystal symmetry.}
\end{figure*}

In this study, we use a comprehensive set of experimental techniques to overcome the limitations in determining the nature of  the interactions governing the ground state of the prototypical quasi-two-dimensional (2D) square-lattice iridate \sio.
It is now well established that a ($\pi$,$\pi$) staggered arrangement of pseudospins---also known as $J_{\textrm{eff}}$=1/2 moments~\cite{Kim2008}---is stabilized by the strong antiferromagnetic (AF) NN Heisenberg interaction ($J$$\sim$\,60 meV~\cite{Kim2012}). This state, which remains intact even when charge carriers are introduced by chemical~\cite{GretarssonRIXS,Pincini} or photo-doping~\cite{Dean2016} to disrupt the static long-range order, underlies a striking parallel between the phenomenology of electron-doped \sio\, and hole-doped cuprates; namely, high-temperature pseudogaps and low-temperature $d$-wave gaps in the single-particle removal spectra~\cite{Kim2014,Kim2015,Yan}. The complex static long-range order (Fig.~\ref{fig1}), that sets in at $T_N$$\simeq$\,230 K~\cite{Crawford} reveals additional interactions at play, including anisotropic interactions that confine the pseudospins to the $ab$-plane, Dzyaloshinsky-Moriya (DM) interactions that cant the pseudospins and add up to a non-zero net moment in each IrO$_2$ layer~\cite{Jackeli}, and interlayer couplings that stabilize the `up-up-down-down' ($uudd$) stacking pattern of the net moments along the $c$-axis~\cite{Kim2009,Dhital,Ye} (see Fig.~\ref{fig:mag_stacking}). 

These interactions manifest as a deviation from the Heisenberg universality class evidenced by the temperature dependence of the order parameter in diffuse x-ray scattering~\cite{Fujiyama,Vale}, a resonance line in electrons spin resonance~\cite{Bahr}, and a spin-wave gap in Raman scattering~\cite{Gim,Gretarsson2017} and RIXS~\cite{Pincini}. However, interpretations of these experiments have led to mutually inconsistent results, and a coherent understanding of the low-energy pseudospin dynamics is still lacking. For instance, the energy scale for the out-of-plane spin-wave gap, a direct measure of the magnetic anisotropy, inferred from these measurements varies widely between $\lesssim$ 1 meV~\cite{Bahr} and 30 meV~\cite{Pincini}. The lack of knowledge about the hierarchy among these interactions is an impediment to our understanding of the mechanism that stabilizes the observed static magnetic structure; notably, the fact that the magnetic easy axis points away from the NN bonds [Fig.~\ref{fig1}(c)]. 

In this paper, we present a comprehensive study of the low-energy pseudospin interactions that generate the static magnetic structure and govern linear responses to magnetic fields of the archetypal spin-orbit Mott insulator \sio. Our work establishes a minimal Hamiltonian that captures the full 3D static magnetic structure and uncovers the essential role of pseudospin-lattice coupling thus far overlooked in most theories of magnetism in strongly spin-orbit coupled materials. Our work has  important implications for all experiments involving quantities that depend on the global symmetry and/or topology of the system, such as the selection rules for the second harmonic generation~\cite{DiMatteo,Torchinsky,Zhao}. Further, it raises the question of the role of lattice degrees of freedom in emergent phases of the square-lattice iridates~\cite{Torchinsky,Zhao,Kim2014,Kim2015,Yan}, which can serve as a model system for electron-lattice interactions in many other correlated electron materials such as the colossal magnetoresistive manganites~\cite{Millis} and high-temperature superconducting cuprates~\cite{Gunnarsson2008}.

\section{Ground state and low field magnetism}

\begin{figure}
\includegraphics[width=1\linewidth]{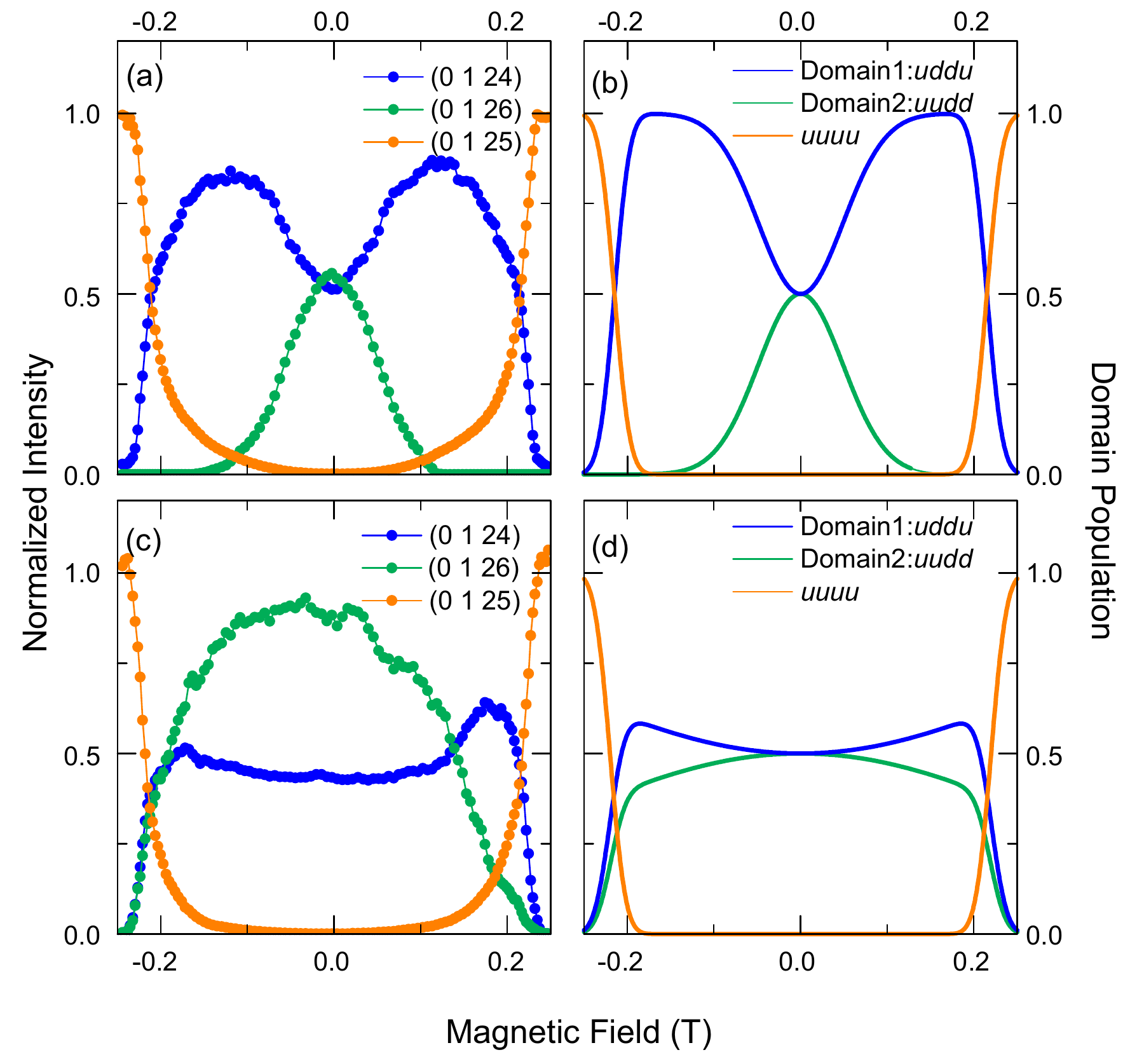}
\caption{\label{fig:xdiff} Normalized RMXS intensity of magnetic reflections (0\,1\,24), (0,1,25) and (0\,1\,26) as a function of magnetic field applied along (a) [0\,1\,0] and (c) [1\,1\,0] compared to simulated domain populations with field applied along (b) [0\,1\,0] and (d) [1\,1\,0]. The data was taken at T=60K, and the intensity has been corrected for structure and polarization factors. At around 0.2 T the stacking pattern changes to $uuuu$ (see Fig.~\ref{fig:mag_stacking}).
}
\end{figure}

\subsection{Magnetic domains}
\begin{figure*}
\includegraphics[width=0.8\linewidth]{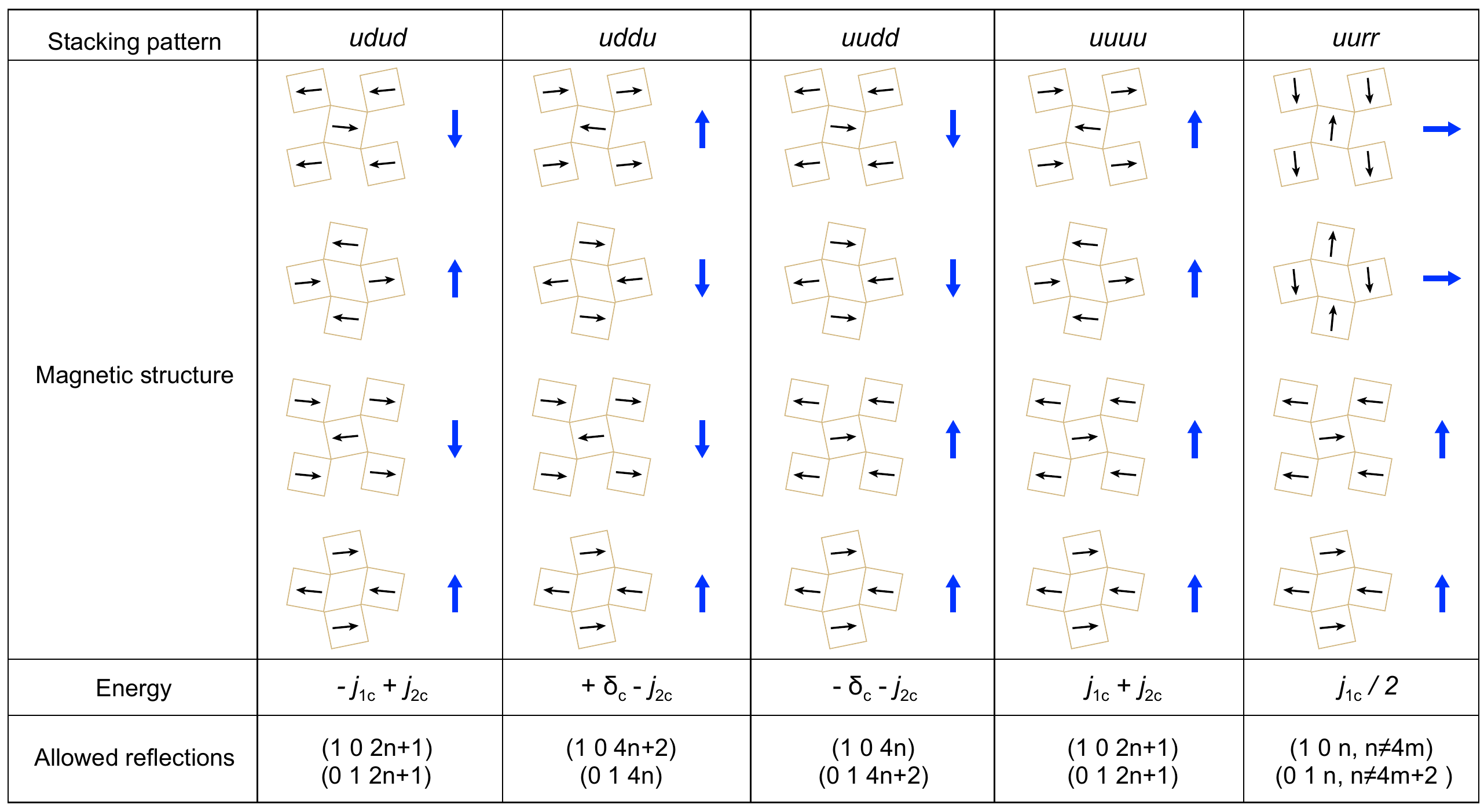}
\caption{\label{fig:mag_stacking}Top view of possible stacking of pseudospins (black arrows) and the corresponding net ferromagnetic moment (blue arrows) in each layer, where the labeling up(\textit{u}), down(\textit{d}), left(\textit{l}) and right(\textit{r}) refers to their orientation in the $ab$ plane. The energy difference between each of these configurations and the ground state of Eq.~(\ref{eqn1}),
and its allowed reflections are indicated at the bottom. The energy is written in terms of the effective couplings between net moments $j_{1c}$=$4S^2J_{1c}\sin^2\phi$, $j_{2c}$=$-S^2J_{2c} (\cos^2\phi-\sin^2\phi)$ and $\delta_c=4S^2\Delta_c\cos^2\phi$, where $\phi$ is the canting angle. \textit{uudd} or \textit{uddu} is stabilized when $j_\mathrm{2c}>0$ and $\left| j_\mathrm{1c}\right| < 2j_\mathrm{2c}$.}
\end{figure*}

We start by discussing all possible magnetic domain configurations and their evolution in magnetic fields in order to disentangle the response from a single domain. The magnetic ordering breaks the 4$_1$ screw axis symmetry of the the crystal structure of \sio\, (four-fold rotation about the $c$-axis followed by the one-quarter translation along the same lattice vector), which means that successive 4$_1$ operations generate four possible magnetic domains. Only two of these can be distinguished macroscopically as the other two are different only by up$\leftrightarrow$down sublattice switching of the N\'eel order. Thus, there are two distinguishable domains: one with the pseudospins mostly along the $b$-axis with $uddu$ stacking of the canted ferromagnetic component, and the other along the $a$-axis with $uudd$ stacking (see Figs.~\ref{fig1}(c) and \ref{fig:mag_stacking}). As discussed later, the correlation between the pseudospin direction and the stacking pattern necessitates inclusion of an anisotropic interlayer coupling [Fig.~\ref{fig:mag_aniso}(b)], which is symmetry allowed and should be generally non-zero. 
 
In our resonant magnetic x-ray scattering (RMXS) experiments, shown in Fig.~\ref{fig:xdiff} (a) and (c), the two domains are visible as two distinct refections, (0\,1\,24) and (0\,1\,26), owing to their two different stacking patterns. When corrected by geometrical and polarization factors, the intensities of the two reflections directly measure the population of the two domains, which we follow as a function of applied magnetic field. The results agree reasonably well with a simulation assuming 50-50 domain population shown in Fig.~\ref{fig:xdiff} (b) and (d).  

With increasing magnetic field applied along the [0\,1\,0] direction [Fig.~\ref{fig:xdiff}(a-b)] the domain with pseudospins along [1\,0\,0] shrinks as the domain with pseudospins along [0\,1\,0] grows. This can be simply understood, since there is a Zeeman energy gain from the net ferromagnetic moments induced along the field giving rise to the (0\,1\,25) reflection, but the domain repopulation involves complex domain wall motions reflected as deviations from linear behavior and hysteresis in the magnetization measurements shown later [Fig.~\ref{fig:magnetization}(b)]. At $\simeq$\,0.1 T, the magnetic domains are fully aligned as can be seen from the vanishing intensity of (0\,1\,26) and the saturation of (0\,1\,24). For fields $H>0.2$ T, the intensity of the (0\,1\,25) reflection probing \textit{uuuu} stacking (Fig.~\ref{fig:mag_stacking}) greatly increases while (0\,1\,24) decreases, indicating a metamagnetic transition where the ferromagnetic moments align with the field. 

When the field is applied along [1\,1\,0] [Fig.\ref{fig:xdiff}(c-d)] both domains remain populated at $\simeq$\,0.1 T as the field has no preference for either of the two domains. Any slight misalignment of the field from the [1\,1\,0] direction leads to an imbalance in the domain population as can be seen from the small difference in the field dependence of the (0\,1\,24) and (0\,1\,26) reflections. The persistence of both domains above 0.1 T implies an anisotropy within the $ab$ plane; without it the ferromagnetic moments would simply rotate perpendicular to the field.  

\subsection{In-plane magnetic anisotropy}

We investigate the in-plane magnetic anisotropy by performing longitudinal magnetization measurements. 

\begin{figure}
\includegraphics[width=0.85\linewidth]{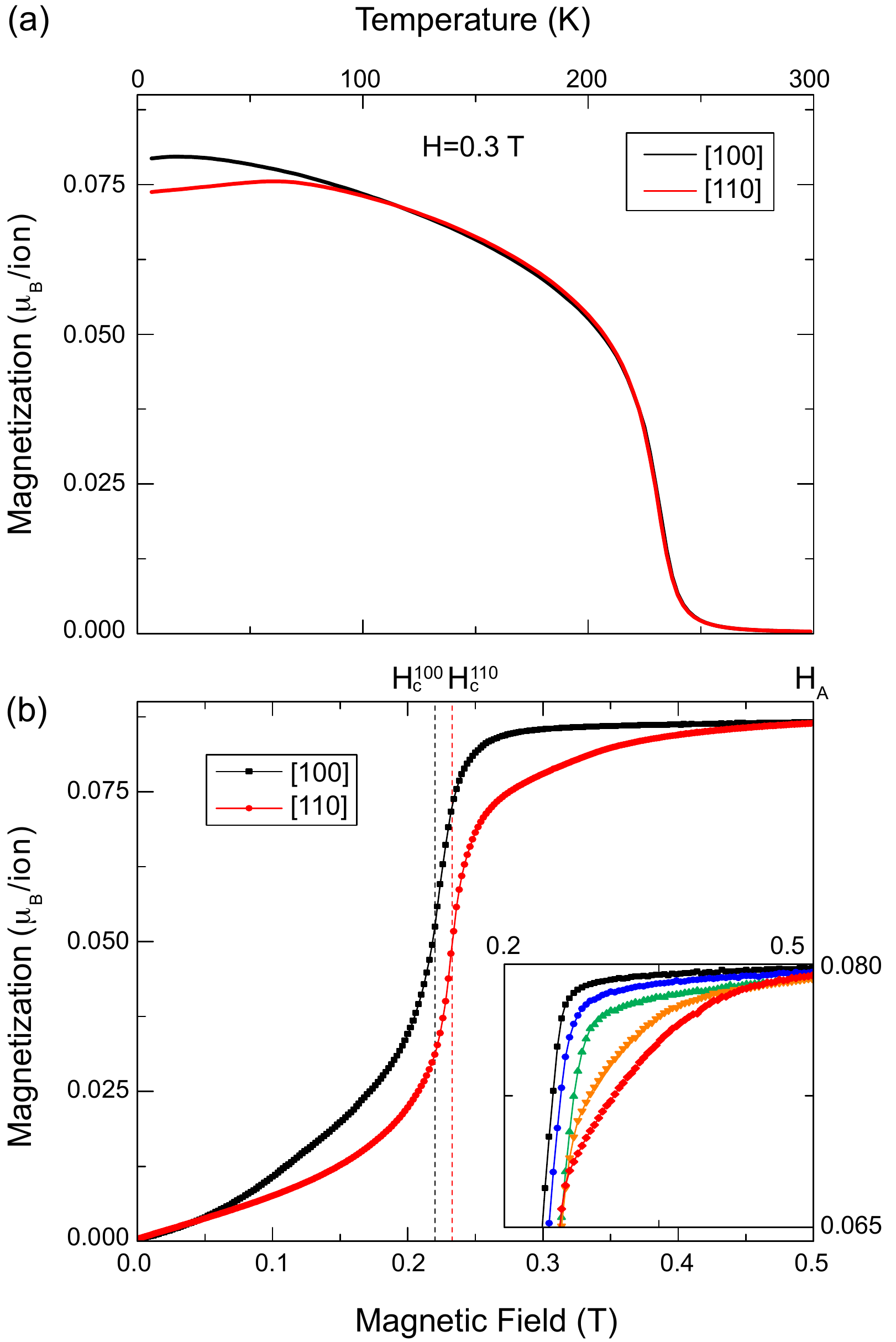}
\caption{\label{fig:magnetization}(a) Magnetization as a function of temperature along  [1\,0\,0] (black) and [1\,1\,0] (red), taken at H = 0.3 T. (b) Magnetization as a function of magnetic field for [1\,0\,0] (black) and [1\,1\,0] (red), taken at T = 5 K.  Inset: detailed measurements for different field angles from [1\,0\,0] to [1\,1\,0] every $11.25^\circ$, focusing in the region where the biggest effect due to anisotropy is seen.}
\end{figure}

 Figure \ref{fig:magnetization}(a) shows the magnetization as a function of temperature with a magnetic field applied along [1\,0\,0] and [1\,1\,0]. The magnetization shows an upturn at T$_\mathrm{N}\simeq230$ K where the system becomes antiferromagnetic. The black curve for measurements along [1\,0\,0] shows an order-parameter-like increase that persists to the lowest temperature, which is characteristic of the weak ferromagnetism, as the applied field of $H$=\,0.3 T is enough to fully align the net moments along [1\,0\,0] [Fig.~\ref{fig:xdiff}(b)]. For the measurements along [1\,1\,0] shown in red, a decrease in the magnetization is observed at low temperature, which points to a temperature dependent anisotropy. We note that it requires a very high quality sample to observe the in-plane anisotropy as it was not visible in previous magnetization measurements~~\cite{Takayama} [see Supplementary Materials (SM) for a description of our samples].

In order to understand the origin of such anisotropy, the magnetization as a function of magnetic field was studied at $T$=\,5 K. In Fig.~\ref{fig:magnetization}(b), anisotropic behavior below 0.5 T is clearly seen. In particular, (i) the two curves for magnetic field along [1\,0\,0] and [1\,1\,0] have different slopes below 0.2 T; (ii) a metamagnetic transition occurs at $H_c^{100}$=\,0.22 T and $H_c^{110}$=\,0.23 T respectively; and (iii) saturation in the magnetization is attained slightly above $H_c^{100}$ along [1\,0\,0] but only at a higher field $H_A\simeq0.5$ T along [1\,1\,0]. The inset shows measurements at different angles between these two limits, showing the gradual change from one behavior to the other. 

For a quantitative analysis of the magnetization measurements, it is necessary to consider the possible mechanisms for in-plane anisotropy.

\subsection{Mechanisms for in-plane anisotropy}

As previously discussed, the Hamiltonian for magnetic interactions in \sio\, is dominated by Heisenberg interactions:

\begin{equation}
H_\mathrm{iso}=\sum_{\langle ij\rangle}J_{ij} \vec{S}_i \cdot \vec{S}_j +J_\mathrm{1c}\vec{S}_i\cdot\vec{S}_j +J_\mathrm{2c}\vec{S}_i \cdot\vec{S}_{j},
\label{eqn1}
\end{equation}
\noindent
where $\vec{S}_i$ labels the pseudospin at site $i$, and $J_{ij}$ denote first $J$, second $J_2$ and third $J_3$ in-plane nearest-neighbor interactions~\cite{Kim2012}. Similarly, $J_\mathrm{1c}$ and $J_\mathrm{2c}$ are the first and second nearest-interlayer interactions as shown in Fig.~\ref{fig1}(b). The nearest layer term $J_\mathrm{1c}$ is partially frustrated due to the staggering of pseudospins in adjacent layers as has been pointed out in an earlier study~\cite{Takayama}. The next nearest layer term $J_\mathrm{2c}$ is responsible for the \textit{uudd} or \textit{uddu} stacking patterns (Fig.~\ref{fig:mag_stacking}).

Additionally, tetragonal distortion and rotation of octahedra lead to symmetric and anti-symmetric exchange anisotropy terms of the form:
\begin{equation}
H_\mathrm{ani}^{(1)}=\sum_{\langle ij\rangle}J_zS_i^zS_j^z +\vec{D}\cdot\left(\vec{S}_i\times\vec{S}_j\right),
\label{eqn2}
\end{equation}
\noindent
where $\vec{D}$ is the DM vector along the $c$-axis and gives rise to the canting angle $\phi$. The Hamiltonian (2) has been discussed in detail~\cite{Cher05} in the context of K$_2$V$_3$O$_8$~\cite{Lums01}. In \sio, these anisotropy terms confine the pseudospins to the $ab$ plane and give rise to an out-of-plane magnon gap~\cite{Jackeli}.

\begin{figure}
\includegraphics[width=1\linewidth]{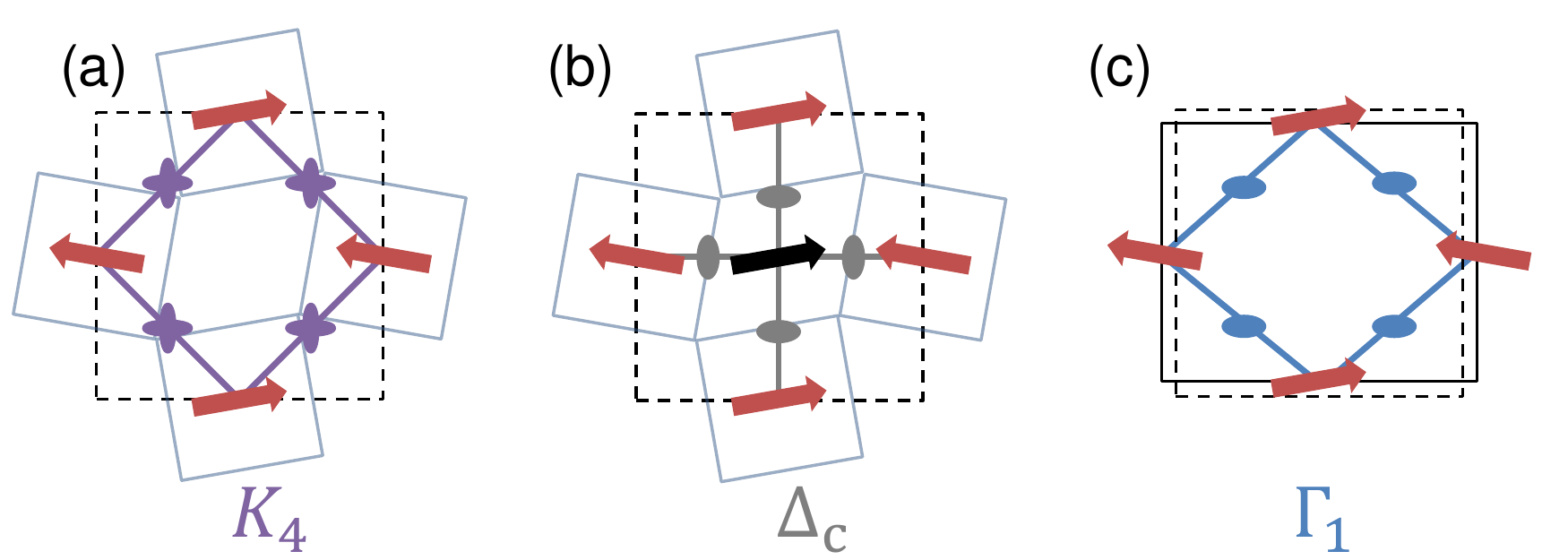}
\caption{\label{fig:mag_aniso}An illustration of possible mechanisms for anisotropy. (a) bi-axial in-plane anisotropy $K_4$ shown as purple ellipses, (b) anisotropy in the out-of-plane nearest neighbor coupling (grey ellipses) connecting pseudospins in two neighboring layers (red and black) and (c) anisotropy $\Gamma_1$ (blue ellipses) due to coupling of the pseudospins to the orthorhombically deformed lattice.}
\end{figure}

The anisotropy within the $ab$-plane is naturally expected as a square lattice has only a discrete, four-fold rotation symmetry. Indeed, it has been observed in recent magnetoresistance~\cite{Wang} and torque magnetometry~\cite{Nauman,Fruchter} measurements. In the latter, a phenomenological biaxial anisotropy energy with magnetic easy axes along the crystallographic $a$ or $b$ axes of the form
\begin{equation}
-K_4 \cos 4\theta
\label{eqn3}
\end{equation}
was considered and is depicted in Fig.~\ref{fig:mag_aniso}(a). $\theta$ is the angle between the canted ferromagnetic moments and $a$. Theoretically, biaxial anisotropy is attained when considering quantum order-by-disorder effects~\cite{Yildirim,Katukuri}.

Another contribution to anisotropy comes from the anisotropic interlayer interaction~\cite{Katukuri}. This can be written as a $4_1$ symmetry allowed Hamiltonian:
\begin{equation}
H_\mathrm{ani}^{(2)}=\sum_{\langle ij\rangle}\pm\Delta_c\left(S_i^aS_j^a-S_i^bS_j^b\right)
\label{eqn4}
\end{equation}
\noindent
where $\left<ij\right>$ run over first nearest-neighbors in adjacent layers and $H_\mathrm{ani}^{(2)}$ changes sign depending on the direction of the bond [see Fig.~\ref{fig:mag_aniso}(b)]. This term lifts the degeneracy between \textit{uudd} and \textit{uddu}  and accounts for the observed magnetic structure: for the domain with the pseudospins mostly along $b$-axis, \textit{uddu} stacking is favored, whereas for the domain with the pseudospins mostly along $a$-axis, \textit{uudd} stacking is preferred [Fig.~\ref{fig1}(c)]. 

In the model put forward by recent theoretical work~\cite{Liu}, the coupling of the pseudospins to the lattice is responsible for the alignment of the moments along the crystallographic $a$ or $b$ directions, and gives rise to in-plane anisotropy, as depicted in Fig.~\ref{fig:mag_aniso}(c). It takes the form:
\begin{eqnarray}
 H_\mathrm{sp-lat}=\sum_{\langle ij\rangle}&& \Gamma_1 \cos{ 2\theta} \left(S_i^xS_j^y+S_i^yS_j^x \right)\nonumber\\
&-&\Gamma_2 \sin{ 2\theta} \left(S_i^xS_j^x-S_i^yS_j^y \right),
\label{eqn5}
\end{eqnarray}
where $x$ and $y$ denote the directions along the Ir-O bonds, $\Gamma_1$ and $\Gamma_2$ are the energy scales of the pseudospin-lattice coupling to distortions along [1\,0\,0] and [1\,1\,0] respectively, scaled by the elasticity parameters and the square of the ordered moment. We note that while $H_\mathrm{sp-lat}$ preserves the four-fold symmetry  per se, it leads to an orthorhombic distortion below $T_N$, and thus generates a uniaxial two-fold anisotropy~\cite{Liu}. A special feature of this model is that the magnetic anisotropy potential is a function of the moment direction itself [via angle $\theta$ in Eq.~(\ref{eqn5})]. This is markedly different from the conventional, constant anisotropy terms $K_4$ and $\Delta_c$ discussed above.

\begin{figure*}
\includegraphics[width=0.7\linewidth]{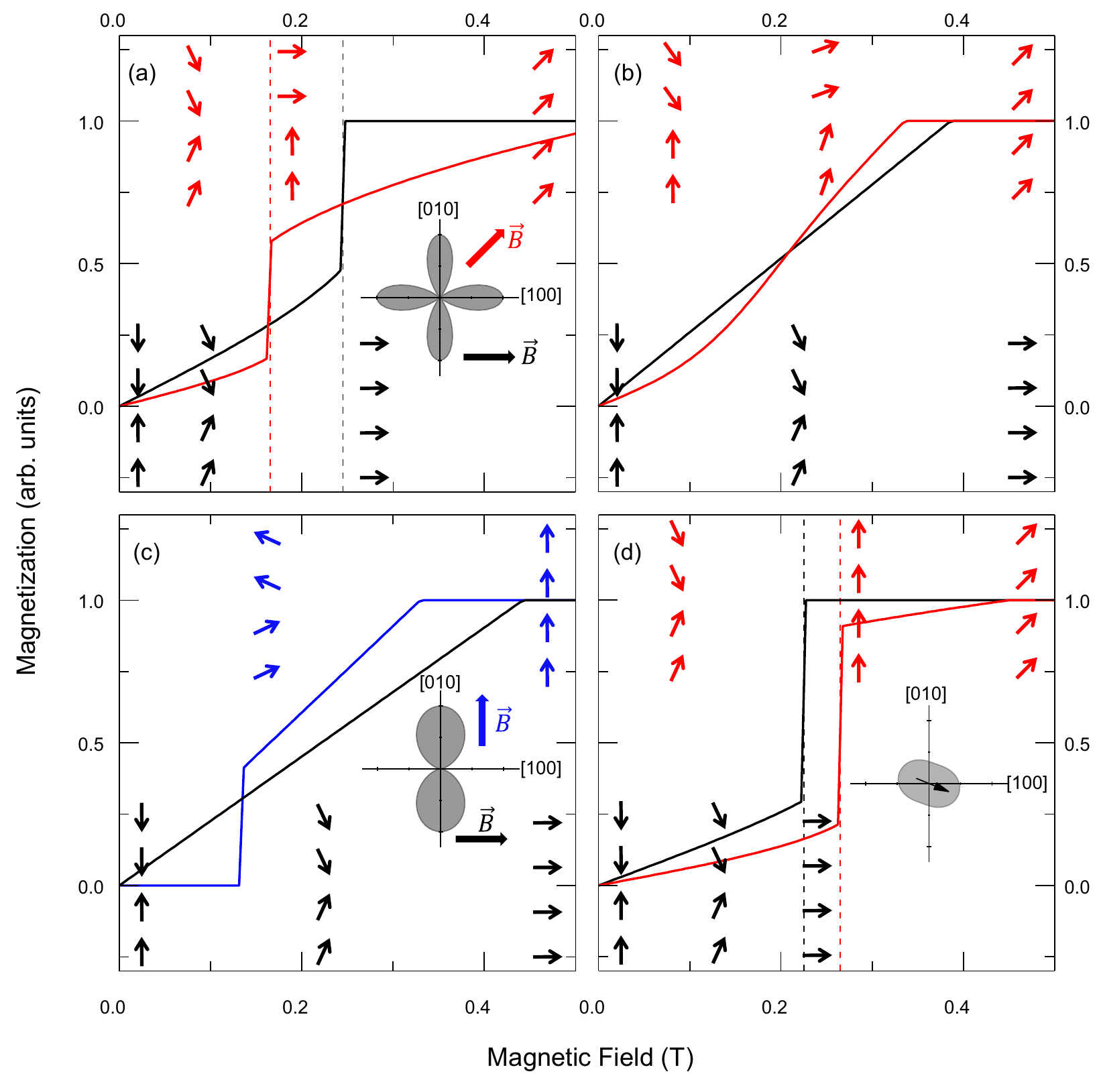}
\caption{\label{fig:MH_model}Model calculation of the magnetization as a function of magnetic field applied along [1\,0\,0] (black) and [1\,1\,0] (red) for (a) biaxial anisotropy $K_4$, (b) anisotropy in the interlayer coupling $\Delta_c$, (d) anisotropy due to spin-lattice coupling $\Gamma_1$, and (c) uniaxial anisotropy along the hard [1\,0\,0] (black) and easy [0\,1\,0] (blue) axes. The parameters used are (a-d) $J_\mathrm{1c}=16.4\ \mu$eV, $J_\mathrm{2c}=-6.2\ \mu$eV, (a,b,d) $\Delta_c=0.02 J_\mathrm{1c}$ , (a) $K_4=2.7\ \mu$eV, (c) $K_2=2.7\ \mu$eV and (d) $\Gamma_1=2.7\ \mu$eV, $\Gamma_2=0$. The moment orientation for different field configurations are shown as colored arrows. The inset of (a), (c) and (d) show schematically the in-plane anisotropy energy. Note that in (d) the anisotropy rotates as the moment does.}
\end{figure*}

We have calculated the ground state configuration in an applied magnetic field and magnetization curves for the anisotropic Hamiltonians discussed above. Figure \ref{fig:MH_model} shows the results for (a) a biaxial anisotropy [Eq.~(\ref{eqn3})], (b) an anisotropy in the interlayer coupling [Eq.~(\ref{eqn4})], (c) a phenomenologial uniaxial anisotropy of the form $-K_2\cos 2\theta$, and (d) an anisotropy due to spin-lattice coupling following Eq.~(\ref{eqn5}), for a set of parameters (indicated in the caption) that best matches the data at $T$=\,5 K. These different types of anisotropies are schematically shown in the inset of each figure.

In Fig.~\ref{fig:MH_model}(a), we consider the case for bi-axial anisotropy. When the field is applied along the $a$-axis, in the favorable domain having moments along the $b$-axis as discussed above, the magnetization increases almost linearly at first, followed by a sudden jump to saturation at some critical field, as the net moments snap to the $a$-axis by the biaxial anisotropy. For field applied along the [1\,1\,0] direction, it takes much higher field to saturate the magnetization as the field has to overcome the biaxial anisotropy. However, it is important to note that a jump in the magnetization occurs at a lower field along the hard axis because there is an intermediate $uurr$ phase (Fig.~\ref{fig:mag_stacking}) that gains more than one-half of the saturation Zeeman energy but does not cost any biaxial anisotropy energy. This is a key feature of the biaxial anisotropy model that differentiates it from the pseudospin-lattice coupling model. The former fails to correctly describe the data in Fig.~\ref{fig:magnetization}(b) as $H_c^{110}$$<$\,$H_c^{100}$ for any set of values of the parameters.

In Fig.~\ref{fig:MH_model}(b), the bare effect of the anisotropy in the out-of-plane coupling on the magnetization measurements is shown. When the field is applied along [1\,0\,0], the net moments cant toward the field resulting in a linear increase of the total magnetization until the net moments are fully aligned along the field. For field along [1\,1\,0], the magnetization deviates from linear behavior, for small fields this direction is harder but becomes easier at some intermediate field compared to [1\,0\,0]. Although this type of anisotropy does not describe our magnetization data at low temperatures, it becomes relevant at higher temperatures as shown in the next section.

For an illustration, we consider in Fig.~\ref{fig:MH_model}(c) a hypothetical situation where the net moments are stabilized along the $b$-axis by a single-axis anisotropy as one would expect if the tetragonal symmetry is reduced to orthorhombic for instance via uniaxial strain. When the field is applied along the hard $a$-axis, the magnetization is again linear. Instead, when the field is applied along the easy $b$-axis, the net moments remain in their zero-field orientation up to some field before they flop perpendicular to the field and then cant toward the field. This happens when the Zeeman energy gain overcomes the single-axis anisotropy. Note that the field along the $b$-axis required to saturate the magnetization is lower than that for the field along the $a$-axis  because the single-axis anisotropy helps alignment along the $b$-axis.

Finally, we consider the pseudospin-lattice coupling model in Fig.~\ref{fig:MH_model}(d). It is clear that it captures the salient features of the data, in particular the fact that $H_c^{100}$$<$\,$H_c^{110}$. As in the case of the biaxial anisotropy model when the field is applied along the $a$-axis, the magnetization jumps to saturation because both $a$-axis and $b$-axis are equally energetically favorable. However, once the moment is aligned along a certain direction and the lattice is distorted along that direction, the pseudospin-lattice coupling effectively acts like a single-axis anisotropy, which means that \textit{uurr} type of stacking is never favored. Instead for an applied field along [1\,1\,0], the  moments flop to a \textit{uuuu} stacking pattern along either $a$ or $b$-axis, before rotating toward the field direction.

Now we turn our attention to other experimental evidence for the pseudospin-lattice coupling.
\subsection{Further evidence for pseudospin-lattice coupling}
\begin{figure}
\includegraphics[width=0.9\linewidth]{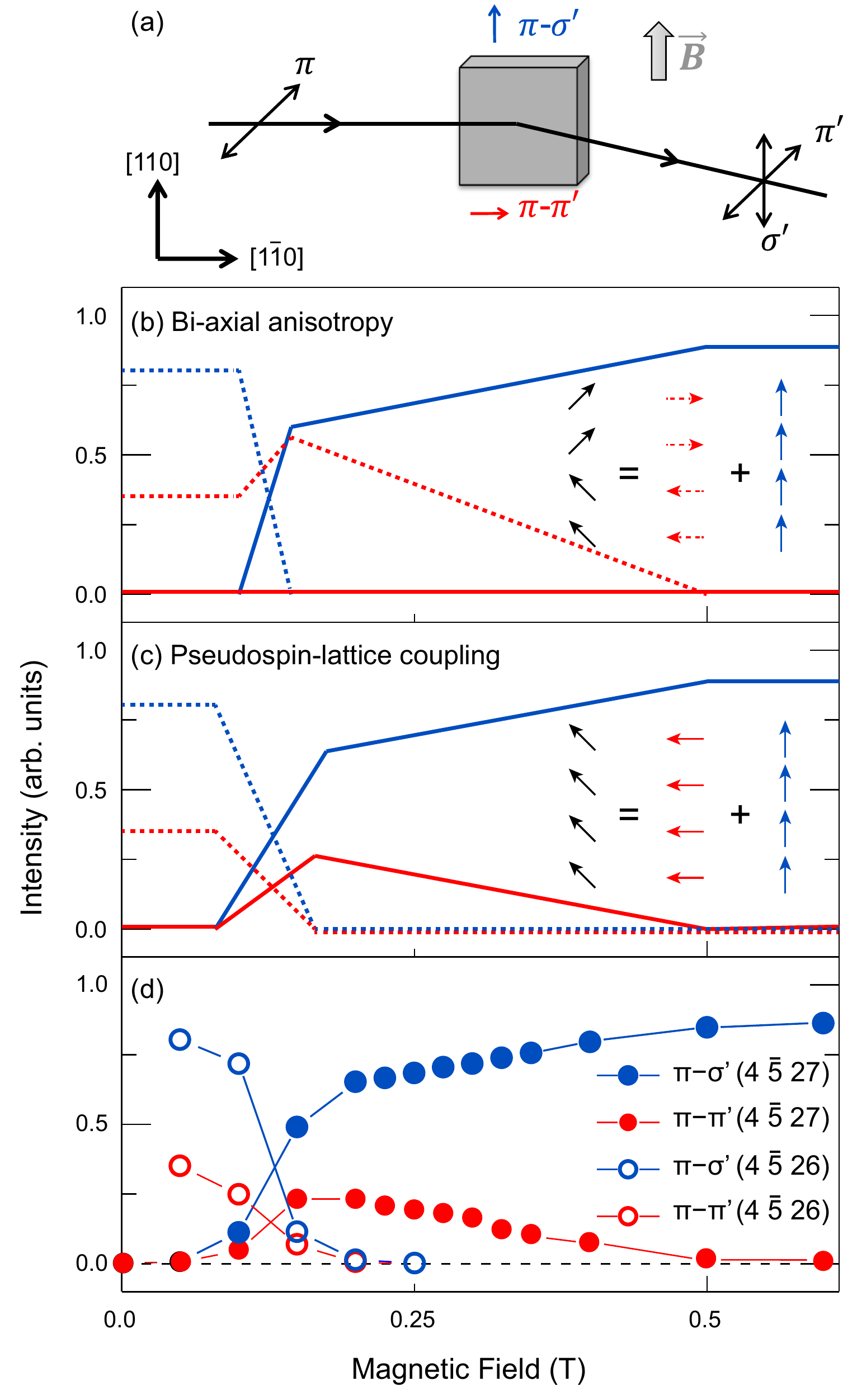}
\caption{\label{fig:REXS}(a) Schematic of the scattering geometry of the RMXS experiment: \psp [\ppp] is sensitive to domains with net moment along [1\,1\,0] ([1\,$\bar{1}$\,0]). Calculated intensity of (4\,$\bar{5}$\,26) and (4\,$\bar{5}$\,27) magnetic reflections for the two polarizations as a function of field for the moment configuration attained with (b)bi-axial anisotropy and (c)anisotropy due to spin-lattice coupling compared to the (d)measured integrated intensity taken at  $T$=\,5K. The inset of (b) and (c) show the characteristic moment configuration in the intermediate field region, that can be separated into two components.}
\end{figure}

\begin{figure}
\includegraphics[width=0.9\linewidth]{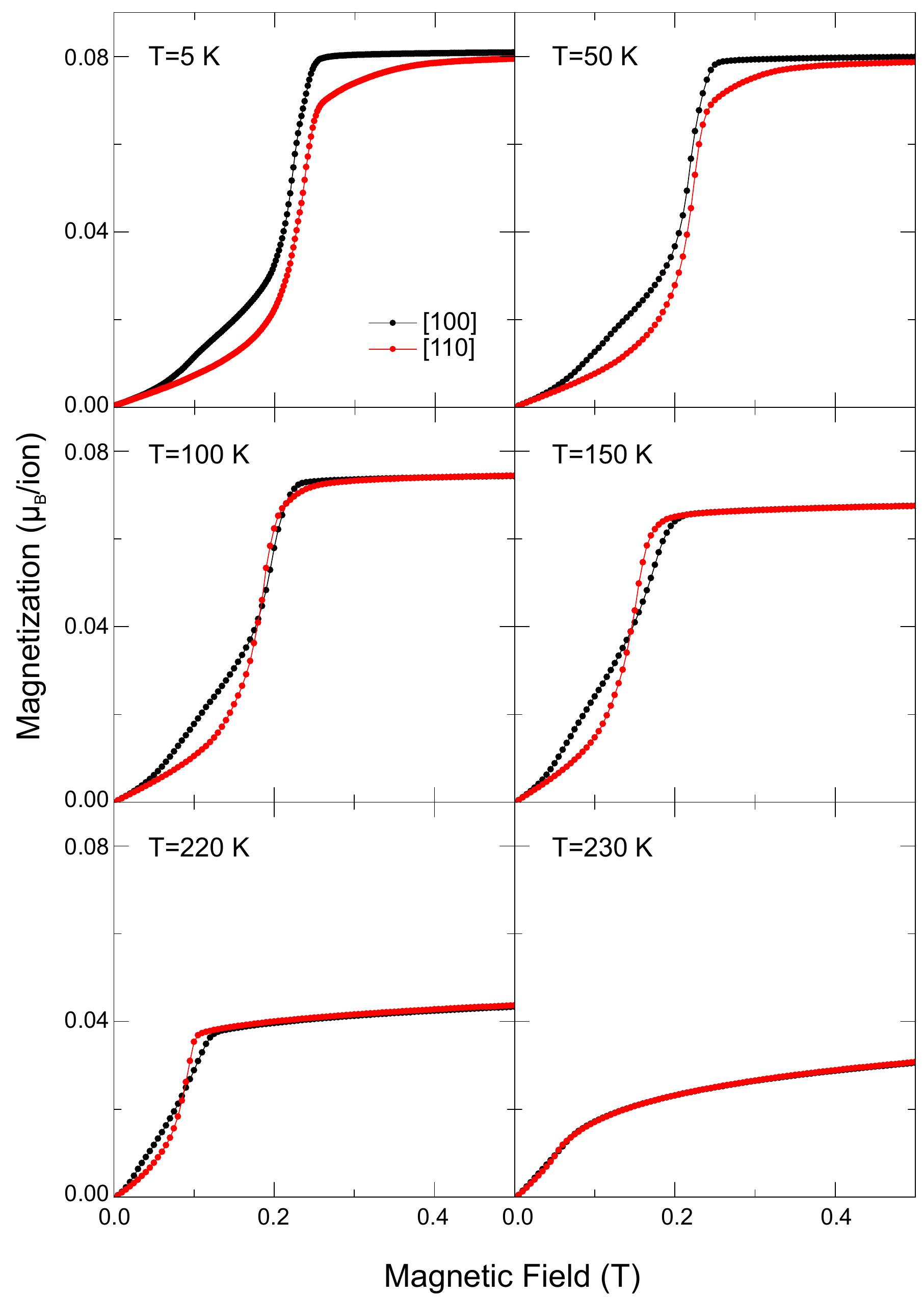}\caption{\label{fig:MH}Temperature evolution of the anisotropy in the magnetization as a function of magnetic field applied along  [1\,0\,0] (black) and [1\,1\,0] (red), taken at  $T$=\,5 K, 50 K, 100 K, 150 K, 220 K and 230 K.}
\end{figure}
To further test the pseudospin-lattice coupling model, we performed RMXS measurements with polarization analysis and a 14 T split-coil cryomagnet at the P09 beamline at DESY (see SM for details). The experimental setup is shown in Fig.~\ref{fig:REXS}(a): $\pi$-polarized x-rays scatter off the \sio\, single crystal with magnetic field applied along the [1\,1\,0] direction. Analysis of the polarization allows to separate magnetic scattering from net moments along [1\,1\,0] \psp and  [1\,$\bar{1}$\,0] \ppp (the main AF components of the pseudospins are opposite).
The \textit{uurr} structure at intermediate fields between $H_c^{110}$ and $H_A$ in the bi-axial anisotropy model can be described as a superposition of \textit{uudd} stacking of moments along [1\,$\bar{1}$\,0] and \textit{uuuu} stacking of moments along [1\,1\,0], giving rise to signals from the (4\,$\bar{5}$\,26) reflection in \ppp and (4\,$\bar{5}$\,27) in \psp respectively as depicted in Fig.~\ref{fig:REXS}(b). In contrast, in the pseudospin-lattice coupling model the moments align along the easy axis with \textit{uuuu} stacking, giving rise to signals from (4\,$\bar{5}$\,27) reflection in both polarization channels, but no signal in (4\,$\bar{5}$\,26), as shown Fig.~\ref{fig:REXS}(c). Our measured data shown in Fig.~\ref{fig:REXS}(d) are in perfect agreement with the pseudospin-lattice coupling model. Together with the magnetization study, these results unambiguously establish that the in-plane anisotropy cannot be explained by pseudospin interactions only, and requires a mechanism that breaks the four-fold symmetry of the underlying lattice, as provided by pseudospin lattice coupling.

Next, we study the temperature dependence of the anisotropy. Figure \ref{fig:MH} shows the magnetization as a function of field measured at various temperatures. As temperature increases both the saturation field $H_A$ and the critical field $H_c^{110}$ decrease, with $H_c^{110}$ becoming smaller than $H_c^{100}$ around 100 K. The anisotropy remains up to $T_\mathrm{N}$, but the characteristic magnetization curves have significantly changed: along [1\,0\,0] the curve is almost linear up to saturation, whereas along [1\,1\,0] a jump at a lower field is still visible. Comparing the data at $T$=\,220 K with the calculated curves for anisotropic interlayer coupling $\Delta_c$ shown in Fig.~\ref{fig:MH_model}(b), we conclude that the temperature dependent in-plane anisotropy vanishes close to $T_\mathrm{N}$, while $\Delta_c$ remains. This is due to the coupling of the pseudospins to the lattice getting largely reduced as the magnetic order disappears~\cite{Liu}. Note that the contribution of $\Delta_c$ does not qualitatively modify the modeled curves for anisotropic magnetization at low temperature and has been included in Fig.~\ref{fig:MH_model}(a,d).

Having established the necessity of both in-plane anisotropy $\Gamma_1$ that dominates magnetic anisotropy at low temperature, and anisotropic interlayer coupling $\Delta_c$, which becomes more important as temperature increases, we turn to their effect on the low energy excitations in \sio.

\section{Magnetic Excitations}
\subsection{Modelling}
\begin{figure}
\includegraphics[width=0.9\linewidth]{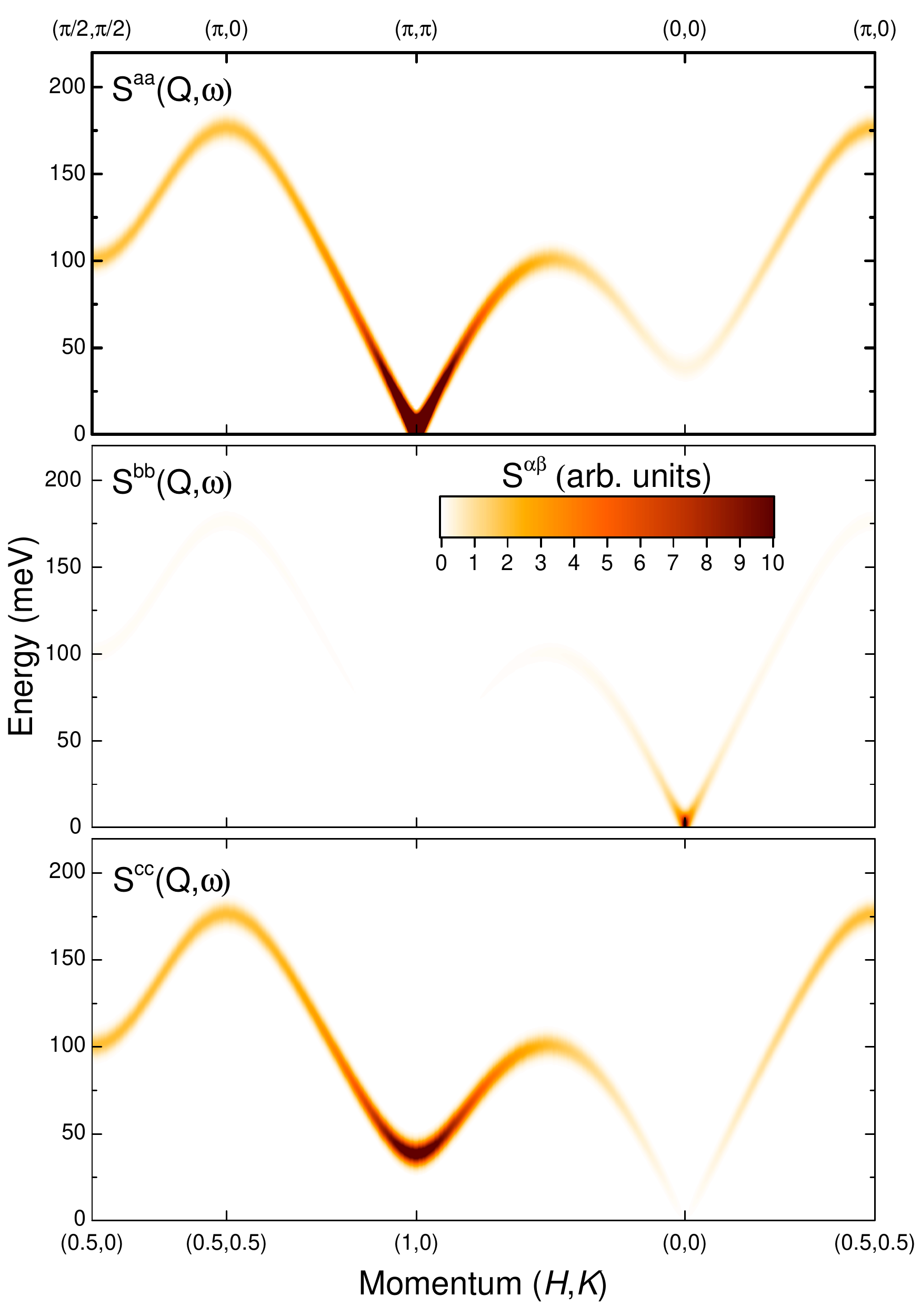}
\caption{\label{fig:disp}Calculated dynamical spin structure factor as a function of momentum $Q$ and energy $\omega$ for spin components along the crystallographic directions $a$ (top), $b$ (middle), and $c$ (bottom). A Gaussian broadening $\delta E$=\,10 meV is used for clarity. The magnetic structure is chosen with the main component of the moments aligned along [1\,0\,0]. The in-plane momenta indicated on the top axis refer to the undistorted square lattice unit cell which is
doubled for the magnetic unit cell indicated in the bottom axis in reciprocal lattice units. The parameters used for the calculation are: $J=57$ meV, $J_2=-16.5$ meV, $J_3=12.4$ meV determined from RIXS measurements~\cite{GretarssonRIXS,Pincini}, $\phi=13^\circ$ determined from neutron diffraction~\cite{Ye}, which gives  $D=28$ meV, $J_z=2.9$ meV. This results in an out-of-plane gap $\Delta _\mathrm{out}=40$ meV consistent with our measurements shown in the SM (but is larger than previously reported~\cite{Pincini}).}
\end{figure}

We model the magnetic excitations using code based on the SpinW library~\cite{spinw}, using the Hamiltonian $H=H_\mathrm{iso}+H_\mathrm{ani}^{(1)}+H_\mathrm{ani}^{(2)}+H_\mathrm{sp-lat}$ from Eqs.~(\ref{eqn1},\ref{eqn2},\ref{eqn4},\ref{eqn5}). The parameters $J_\mathrm{1c}=16.4\ \mu$eV, $J_\mathrm{2c}=-6.2\ \mu$eV, $\Delta_c=0.02 J_\mathrm{1c}$ and $\Gamma_1=2.7\ \mu$eV are fixed from fits to the above magnetization data, and a quantum renormalization factor $Z_c=1.67$ is applied based on recent calculations for the spin-1/2 Heisenberg AF on a square lattice~\cite{Jiang}. The parameters for high-energy terms (known from previous RIXS spectra~\cite{GretarssonRIXS,Pincini}) are  listed in the caption of Fig. \ref{fig:disp}. The ground state has two degenerate solutions: antiferromagnetic moments along [1\,0\,0] with \textit{uudd} stacking, or along [0\,1\,0] with \textit{uddu} stacking. Given that there are 4 atoms per sublattice per unit cell, 8 modes are expected: 4 in-plane and 4 out-of-plane. Figure \ref{fig:disp} shows the magnon dispersions and intensities for different spin components. As expected, the in-plane modes are almost degenerate and have a small gap at (1,0) in the two-dimensional Brillouin zone, whereas the out-of-plane modes also practically degenerate have a larger gap $\Delta_\mathrm{out}\simeq4S\sqrt{2J\left(D \tan\phi-J_z\right)}=\,40$ meV. Note that the in-plane modes are visible in both $S_{aa}$ and $S_{bb}$ due to the canting of the moments.

\begin{figure}
\includegraphics[width=0.8\linewidth]{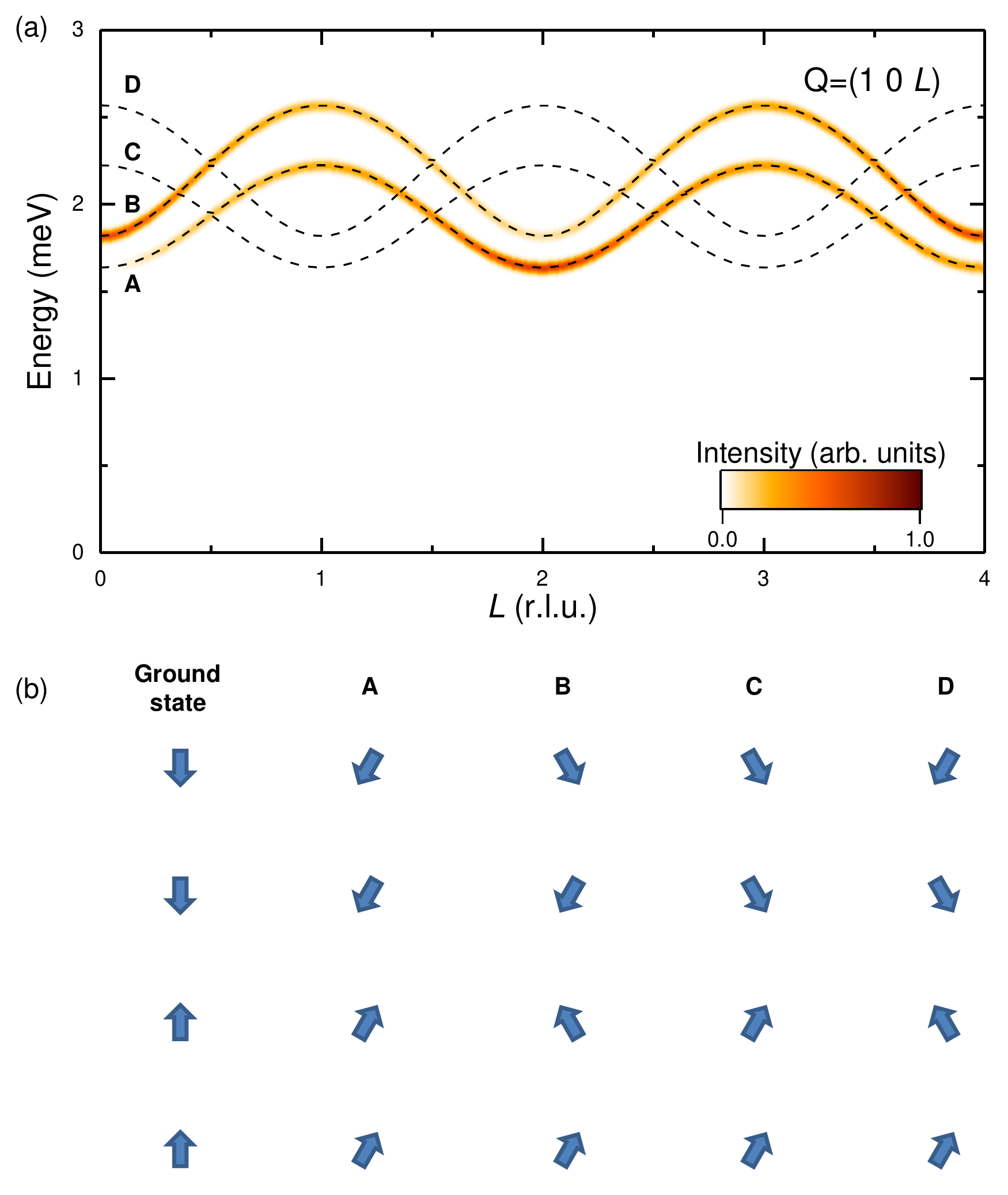}
\caption{\label{fig:ins_twins}(a) Calculated dynamical spin structure factor $S_\mathrm{perp}$ perpendicular to $\vec{Q}$ close to the magnetic zone center as a function of $L$ in reciprocal lattice units and energy, taken into account both magnetic twin domains present in the sample. A $\delta E=0.05$meV Gaussian broadening is used for clarity. (b) Real space representation for $Q=(1\,0\,0)$ of the four different magnon modes (A-D), where blue arrows represent the rotated net moments for each layer. The same parameters as for Fig.\ref{fig:disp} were used.}
\end{figure}

Figure \ref{fig:ins_twins}(a) shows the calculated low-energy excitations close to the magnetic zone center, where the orientation factor for inelastic neutron scattering has been taken into account for ease of comparison with the experiment. This calculation includes both magnetic twin domains present in the sample; however, the scattering from the domain with moments pointing along [1\,0\,0] is largely reduced due to the orientation factor. The splitting due to the effective interlayer coupling $j_{1c}$ of the 4 in-plane modes can be clearly seen in the dispersions, with the bandwidth given by the effective coupling $j_{1c}$+$2j_{2c}$(as defined in the caption of Fig~\ref{fig:mag_stacking}). An increased splitting at $L=2n$ for the upper two branches and at $L=2n+1$ for the lower two, is related to the anisotropy of the interlayer coupling $\Delta_c$. 
This gives a character for each mode related to the stacking patterns of the excited modes as shown in Fig.~\ref{fig:ins_twins}(b) for $Q=(1\,0\,0)$.
Finally, the gap at $Q=(1\,0\,2)$ is due to the in-plane anisotropy $\Gamma_1$.

\subsection{Inelastic neutron scattering}
\begin{figure}
\includegraphics[width=0.95\linewidth]{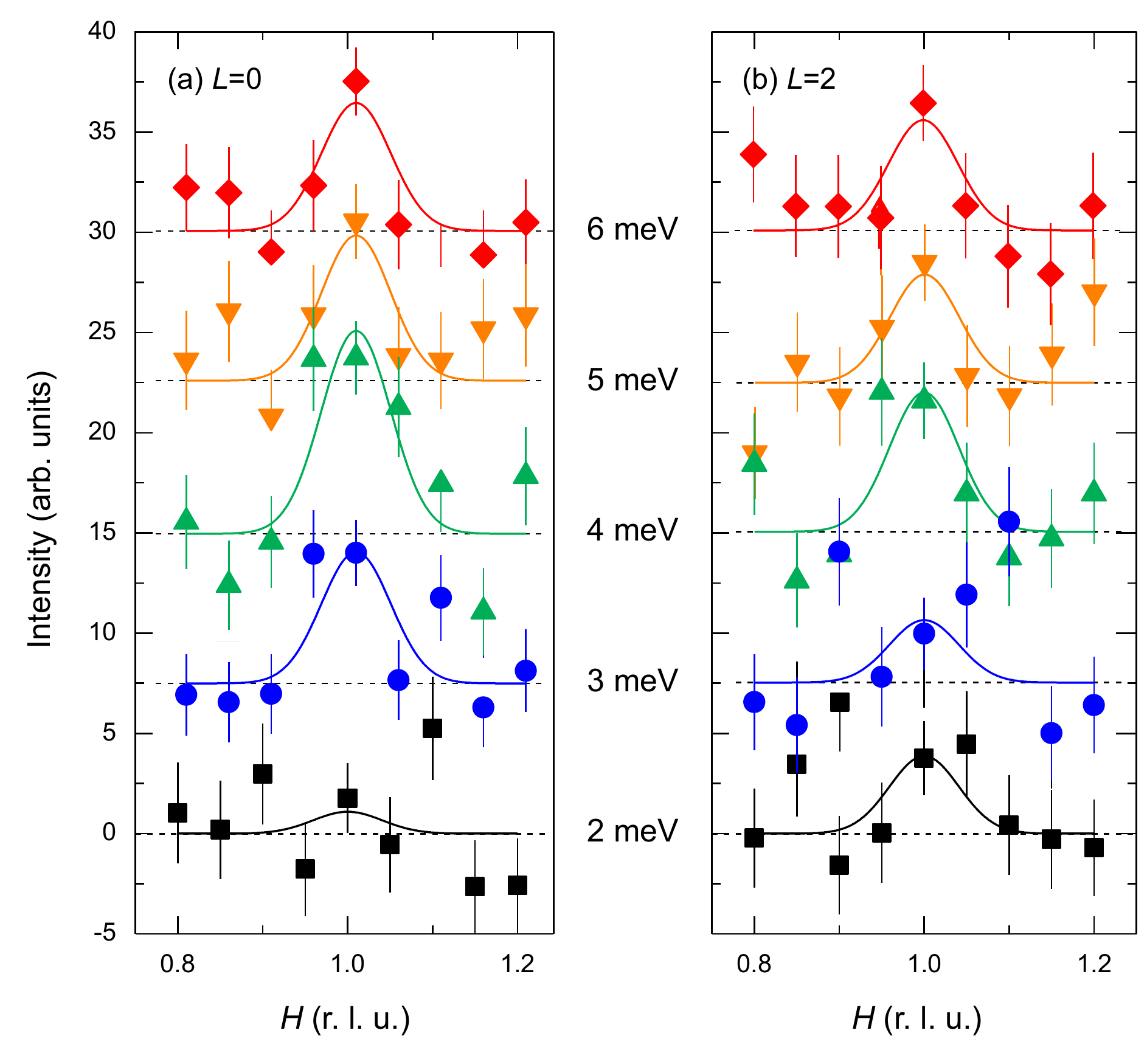}
\caption{\label{fig:INS}Inelastic neutron scattering intensity as a function of $H$ in r.l.u. close to the magnetic zone center (a)(1\,0\,0) and (b)(1\,0\,2), measured for energy transfer $E$ from 2 to 6 meV. The intensity scale is approximately counts per 10 minutes. Lines are results of constrained Gaussian fits with amplitudes and a common width as fitting parameters. A common background has been subtracted from the data, and a vertical offset is used for clarity.}
\end{figure}

Inelastic neutron scattering (INS) is customarily used to study magnons in AF materials, as the neutron cross section for magnetic scattering is sizeable, and sub-meV energy resolution is readily available. In the case of \sio, neutron absorption from Ir nuclei is strong, which, coupled with a relatively small magnetic moment, makes INS measurements challenging. The experiment was conducted using the Three Axis instrument for Low Energy Spectrometry ThALES of the Institut Laue-Langevin. A $\delta_E$=\,0.15 meV resolution and minimal extrinsic background were achieved using PG(002) monochromator and analyzer, and keeping $k_f=1.55\ \text{\AA}^{-1}$ with cold Be as a filter. To maximize the magnon signal, an array of $\sim$300 crystals co-aligned on Al sheets was measured at the magnetic zone centers (1\,0\,0) and (1\,0\,2) where the magnetic form factor and neutron absorption are manageable. In Fig.~\ref{fig:INS}, $H$-scans across the magnetic zone center are shown. Below 2 meV no magnetic signal can be discerned above the background level, whereas a magnon peak emerges at higher energies. This puts an upper bound on the magnon gap $\Delta_\mathrm{INS}=2$ meV. Note that at 2 meV a slightly higher intensity is seen for $Q=(1\,0\,2)$ in comparison to $Q=(1\,0\,0)$, which might be related to seeing the A and B modes respectively(Fig~\ref{fig:ins_twins}).

\subsection{Resonant inelastic x-ray scattering}
\begin{figure}
\includegraphics[width=0.9\linewidth]{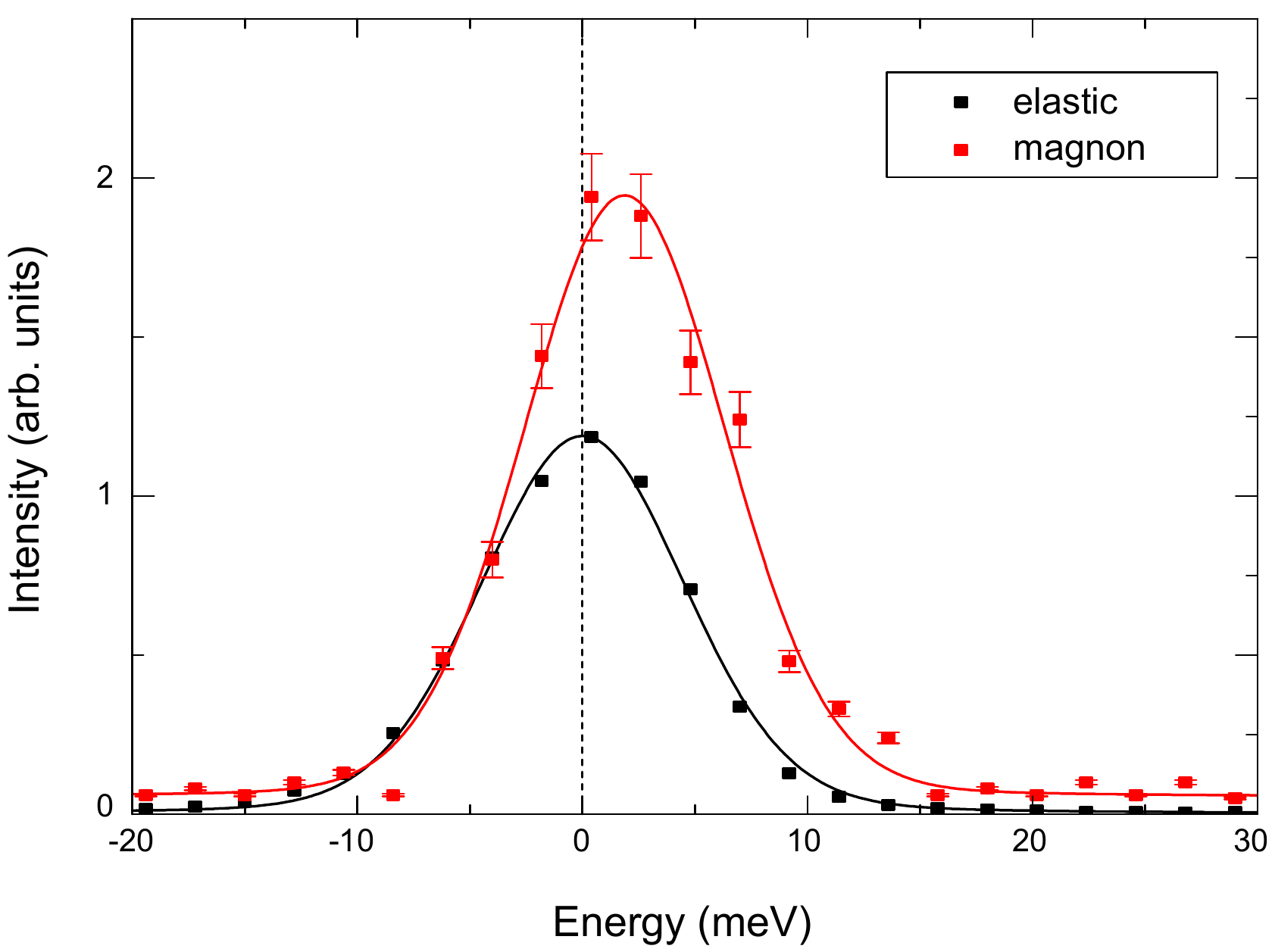}
\caption{\label{fig:RIXS}High-resolution RIXS intensity as a function of energy for (black) incoherent scattering of a scotch tape used as reference, and (red) in-plane magnon mode in \sio\, measured at $Q=$(3\,2\,28.2) close to the magnetic zone center.}
\end{figure}

In order to confirm the in-plane nature of the magnon mode in the INS spectra, we cross-checked the results using high-resolution RIXS measurements at the Ir-$L_3$ edge (E=\,11.215 keV) at the 27-ID of the APS~\cite{Said,jkim2018}. To achieve a 10 meV energy resolution, an incident beam of 11.215 keV was monochromated using a double-crystal diamond high-heat load monochromator and its bandpass was further reduced to 8.9 meV using a four-bounce symmetric Si (844) high-resolution monochromator. The beam was focused to a spot size of 10 $\mu$m$\times$40 $\mu$m FWHM(V$\times$H) on the sample using a KB-focusing mirror system. Scattered radiations from the sample are analyzed by a diced spherical quartz (309), which has an intrinsic bandpass of 3.7 meV at the Ir-$L_3$ edge~\cite{Said}. The in-plane magnon gap was probed at $Q=$(3\,2\,28.2) where the sample surface is at a grazing angle to the incident beam and both \psp and \ppp probe only in-plane magnetic excitations. To have a resolution-limited magnon peak, a high momentum resolution is of particular importance given the relatively high spin-wave velocity; a 3~mm rectangular mask on the analyzer (on a 2~m diameter Rowland circle) was used, giving a maximum $0.086^\circ$ divergence, which translates to $\delta q$=0.0054\ r.l.u. in each in-plane direction and 0.4\ r.l.u. in the out-of-plane direction at $Q=$(3\,2\,28.2). Figure \ref{fig:RIXS} shows the in-plane magnon measured at $Q=$(3\,2\,28.2), from which the in-plane anisotropy gap is estimated to be 2 meV from the peak energy position. This is unequivocally smaller than the out-of-plane magnon gap seen in the standard geometry, that has been previously measured~\cite{Pincini}(see SM for a comparison of our measurements of the two magnon gaps). 

The in-plane gap values measured with RIXS and INS are consistent with the calculated energies from the model, as well as the magnetic excitation emerging below $T_\mathrm{N}$ identified by Raman scattering in previous studies~\cite{Gretarsson2017,Gim}; small differences are due to the interlayer couplings and dispersions along the $L$-direction.

\section{Conclusions}
To conclude, we have unambiguously shown that the equilibrium arrangements of pseudospins in the archetypal spin-orbit Mott insulator \sio\, cannot be explained by considering interactions among pseudospins alone, and that their coupling to the lattice is essential for a quantitative description of the ground state.  

The $uddu$ and $uudd$ stacking patterns of the net ferromagnetic moments peculiar to \sio\, provide a means to differentiate among different types of magnetic anisotropy as they undergo non-trivial changes under moderate applied fields: any four-fold symmetric magnetic anisotropy  necessarily leads to the $uurr$ stacking pattern stabilized in some range of field strength, the absence of which unequivocally implies the reduced symmetry due to magnetostriction. We have directly confirmed that the magnetic structure evolves under applied field as expected in the pseudospin-lattice coupling model by using RMXS.

In the magnetization measurements, the critical fields of the metamagnetic transitions induced by fields applied along the [100] and [110] directions ($H_c^{100}$$<$\,$H_c^{110}$) not only contain information on the symmetry of the magnetic anisotropy, but also allow quantitative extraction of the magnitudes of interlayer couplings and anisotropy parameters. We have shown that the anisotropy of the nearest-neighbor interlayer coupling is responsible for the lifting of the degeneracy of $uddu$ and $uudd$ stacking patterns: for pseudospins along the $a$($b$) axis, $uddu$($uudd$) is stablized. The interlayer coupling anisotropy is most manifest near the N\'eel temperature as the anisotropy due to the pseudospin-lattice coupling becomes suppressed with reduced moment size. 

In the INS and RIXS spectra, the anisotropy due to pseudospin-lattice coupling is largely responsible for the in-plane magnon gap. The measured gap is consistent with our model using the parameters extracted from the magnetometry. 

The two anisotropic interactions uncovered in this study are of particular importance for determining the magnetic ground state of the system, and give a complete description of the magnetism in \sio.

The comprehensive understanding of the magnetic interactions and the magnetoelastic coupling in this archetypical model compound provides a firm basis for the interpretation of thermodynamic and spectroscopic data on other compounds with 4d and 5d valence electrons in various lattice geometries. For instance, the evidence for unconventional order parameters in iridates  with various forms of disorder ~\cite{Zhao} should be critically re-examined in the light of the crucial influence of pseudospin-lattice interactions on the magnetic ground state and excitations in a stoichiometric parent compound. Further, pseudospin-lattice coupling is expected to become of particular importance for the phase behavior of Kitaev-model materials~\cite{Wint17}, where pseudospin frustration leads to a large number of competing many-body states.

\acknowledgments{This work was supported by IBS-R014-A2. 

This research used resources of the Advanced Photon Source, a U.S. Department of Energy (DOE) Office of Science User Facility operated for the DOE Office of Science by Argonne National Laboratory under Contract No. DE-AC02-06CH11357.}

\bibliography{biblio}

\begin{thebibliography}{50}%
\makeatletter
\providecommand \@ifxundefined [1]{%
 \@ifx{#1\undefined}
}%
\providecommand \@ifnum [1]{%
 \ifnum #1\expandafter \@firstoftwo
 \else \expandafter \@secondoftwo
 \fi
}%
\providecommand \@ifx [1]{%
 \ifx #1\expandafter \@firstoftwo
 \else \expandafter \@secondoftwo
 \fi
}%
\providecommand \natexlab [1]{#1}%
\providecommand \enquote  [1]{``#1''}%
\providecommand \bibnamefont  [1]{#1}%
\providecommand \bibfnamefont [1]{#1}%
\providecommand \citenamefont [1]{#1}%
\providecommand \href@noop [0]{\@secondoftwo}%
\providecommand \href [0]{\begingroup \@sanitize@url \@href}%
\providecommand \@href[1]{\@@startlink{#1}\@@href}%
\providecommand \@@href[1]{\endgroup#1\@@endlink}%
\providecommand \@sanitize@url [0]{\catcode `\\12\catcode `\$12\catcode
  `\&12\catcode `\#12\catcode `\^12\catcode `\_12\catcode `\%12\relax}%
\providecommand \@@startlink[1]{}%
\providecommand \@@endlink[0]{}%
\providecommand \url  [0]{\begingroup\@sanitize@url \@url }%
\providecommand \@url [1]{\endgroup\@href {#1}{\urlprefix }}%
\providecommand \urlprefix  [0]{URL }%
\providecommand \Eprint [0]{\href }%
\providecommand \doibase [0]{http://dx.doi.org/}%
\providecommand \selectlanguage [0]{\@gobble}%
\providecommand \bibinfo  [0]{\@secondoftwo}%
\providecommand \bibfield  [0]{\@secondoftwo}%
\providecommand \translation [1]{[#1]}%
\providecommand \BibitemOpen [0]{}%
\providecommand \bibitemStop [0]{}%
\providecommand \bibitemNoStop [0]{.\EOS\space}%
\providecommand \EOS [0]{\spacefactor3000\relax}%
\providecommand \BibitemShut  [1]{\csname bibitem#1\endcsname}%
\let\auto@bib@innerbib\@empty
\bibitem [{\citenamefont {Khaliullin}(2005)}]{Khaliullin2005}%
  \BibitemOpen
  \bibfield  {author} {\bibinfo {author} {\bibfnamefont {G.}~\bibnamefont
  {Khaliullin}},\ }\href {\doibase 10.1143/PTPS.160.155} {\bibfield  {journal}
  {\bibinfo  {journal} {Prog. Theor. Phys. Suppl.}\ }\textbf {\bibinfo {volume}
  {160}},\ \bibinfo {pages} {155} (\bibinfo {year} {2005})}\BibitemShut
  {NoStop}%
\bibitem [{\citenamefont {Jackeli}\ and\ \citenamefont
  {Khaliullin}(2009)}]{Jackeli}%
  \BibitemOpen
  \bibfield  {author} {\bibinfo {author} {\bibfnamefont {G.}~\bibnamefont
  {Jackeli}}\ and\ \bibinfo {author} {\bibfnamefont {G.}~\bibnamefont
  {Khaliullin}},\ }\href {\doibase 10.1103/PhysRevLett.102.017205} {\bibfield
  {journal} {\bibinfo  {journal} {Phys. Rev. Lett.}\ }\textbf {\bibinfo
  {volume} {102}},\ \bibinfo {pages} {017205} (\bibinfo {year}
  {2009})}\BibitemShut {NoStop}%
\bibitem [{\citenamefont {Chaloupka}\ \emph {et~al.}(2010)\citenamefont
  {Chaloupka}, \citenamefont {Jackeli},\ and\ \citenamefont
  {Khaliullin}}]{Chal10}%
  \BibitemOpen
  \bibfield  {author} {\bibinfo {author} {\bibfnamefont {J.}~\bibnamefont
  {Chaloupka}}, \bibinfo {author} {\bibfnamefont {G.}~\bibnamefont {Jackeli}},
  \ and\ \bibinfo {author} {\bibfnamefont {G.}~\bibnamefont {Khaliullin}},\
  }\href {\doibase 10.1103/PhysRevLett.105.027204} {\bibfield  {journal}
  {\bibinfo  {journal} {Phys. Rev. Lett.}\ }\textbf {\bibinfo {volume} {105}},\
  \bibinfo {pages} {027204} (\bibinfo {year} {2010})}\BibitemShut {NoStop}%
\bibitem [{\citenamefont {Choi}\ \emph {et~al.}(2012)\citenamefont {Choi},
  \citenamefont {Coldea}, \citenamefont {Kolmogorov}, \citenamefont
  {Lancaster}, \citenamefont {Mazin}, \citenamefont {Blundell}, \citenamefont
  {Radaelli}, \citenamefont {Singh}, \citenamefont {Gegenwart}, \citenamefont
  {Choi}, \citenamefont {Cheong}, \citenamefont {Baker}, \citenamefont
  {Stock},\ and\ \citenamefont {Taylor}}]{Choi12}%
  \BibitemOpen
  \bibfield  {author} {\bibinfo {author} {\bibfnamefont {S.~K.}\ \bibnamefont
  {Choi}}, \bibinfo {author} {\bibfnamefont {R.}~\bibnamefont {Coldea}},
  \bibinfo {author} {\bibfnamefont {A.~N.}\ \bibnamefont {Kolmogorov}},
  \bibinfo {author} {\bibfnamefont {T.}~\bibnamefont {Lancaster}}, \bibinfo
  {author} {\bibfnamefont {I.~I.}\ \bibnamefont {Mazin}}, \bibinfo {author}
  {\bibfnamefont {S.~J.}\ \bibnamefont {Blundell}}, \bibinfo {author}
  {\bibfnamefont {P.~G.}\ \bibnamefont {Radaelli}}, \bibinfo {author}
  {\bibfnamefont {Y.}~\bibnamefont {Singh}}, \bibinfo {author} {\bibfnamefont
  {P.}~\bibnamefont {Gegenwart}}, \bibinfo {author} {\bibfnamefont {K.~R.}\
  \bibnamefont {Choi}}, \bibinfo {author} {\bibfnamefont {S.-W.}\ \bibnamefont
  {Cheong}}, \bibinfo {author} {\bibfnamefont {P.~J.}\ \bibnamefont {Baker}},
  \bibinfo {author} {\bibfnamefont {C.}~\bibnamefont {Stock}}, \ and\ \bibinfo
  {author} {\bibfnamefont {J.}~\bibnamefont {Taylor}},\ }\href {\doibase
  10.1103/PhysRevLett.108.127204} {\bibfield  {journal} {\bibinfo  {journal}
  {Phys. Rev. Lett.}\ }\textbf {\bibinfo {volume} {108}},\ \bibinfo {pages}
  {127204} (\bibinfo {year} {2012})}\BibitemShut {NoStop}%
\bibitem [{\citenamefont {Chaloupka}\ \emph {et~al.}(2013)\citenamefont
  {Chaloupka}, \citenamefont {Jackeli},\ and\ \citenamefont
  {Khaliullin}}]{Chal13}%
  \BibitemOpen
  \bibfield  {author} {\bibinfo {author} {\bibfnamefont {J.}~\bibnamefont
  {Chaloupka}}, \bibinfo {author} {\bibfnamefont {G.}~\bibnamefont {Jackeli}},
  \ and\ \bibinfo {author} {\bibfnamefont {G.}~\bibnamefont {Khaliullin}},\
  }\href {\doibase 10.1103/PhysRevLett.110.097204} {\bibfield  {journal}
  {\bibinfo  {journal} {Phys. Rev. Lett.}\ }\textbf {\bibinfo {volume} {110}},\
  \bibinfo {pages} {097204} (\bibinfo {year} {2013})}\BibitemShut {NoStop}%
\bibitem [{\citenamefont {Chun}\ \emph {et~al.}(2015)\citenamefont {Chun},
  \citenamefont {Kim}, \citenamefont {Kim}, \citenamefont {Zheng},
  \citenamefont {Stoumpos}, \citenamefont {Malliakas}, \citenamefont
  {Mitchell}, \citenamefont {Mehlawat}, \citenamefont {Singh}, \citenamefont
  {Choi}, \citenamefont {Gog}, \citenamefont {Al-Zein}, \citenamefont {Sala},
  \citenamefont {Krisch}, \citenamefont {Chaloupka}, \citenamefont {Jackeli},
  \citenamefont {Khaliullin},\ and\ \citenamefont {Kim}}]{Hwan15}%
  \BibitemOpen
  \bibfield  {author} {\bibinfo {author} {\bibfnamefont {S.~H.}\ \bibnamefont
  {Chun}}, \bibinfo {author} {\bibfnamefont {J.-W.}\ \bibnamefont {Kim}},
  \bibinfo {author} {\bibfnamefont {J.}~\bibnamefont {Kim}}, \bibinfo {author}
  {\bibfnamefont {H.}~\bibnamefont {Zheng}}, \bibinfo {author} {\bibfnamefont
  {C.~C.}\ \bibnamefont {Stoumpos}}, \bibinfo {author} {\bibfnamefont {C.~D.}\
  \bibnamefont {Malliakas}}, \bibinfo {author} {\bibfnamefont {J.~F.}\
  \bibnamefont {Mitchell}}, \bibinfo {author} {\bibfnamefont {K.}~\bibnamefont
  {Mehlawat}}, \bibinfo {author} {\bibfnamefont {Y.}~\bibnamefont {Singh}},
  \bibinfo {author} {\bibfnamefont {Y.}~\bibnamefont {Choi}}, \bibinfo {author}
  {\bibfnamefont {T.}~\bibnamefont {Gog}}, \bibinfo {author} {\bibfnamefont
  {A.}~\bibnamefont {Al-Zein}}, \bibinfo {author} {\bibfnamefont {M.~M.}\
  \bibnamefont {Sala}}, \bibinfo {author} {\bibfnamefont {M.}~\bibnamefont
  {Krisch}}, \bibinfo {author} {\bibfnamefont {J.}~\bibnamefont {Chaloupka}},
  \bibinfo {author} {\bibfnamefont {G.}~\bibnamefont {Jackeli}}, \bibinfo
  {author} {\bibfnamefont {G.}~\bibnamefont {Khaliullin}}, \ and\ \bibinfo
  {author} {\bibfnamefont {B.~J.}\ \bibnamefont {Kim}},\ }\href {\doibase
  10.1038/nphys3322} {\bibfield  {journal} {\bibinfo  {journal} {Nat. Phys.}\
  }\textbf {\bibinfo {volume} {11}},\ \bibinfo {pages} {462} (\bibinfo {year}
  {2015})}\BibitemShut {NoStop}%
\bibitem [{\citenamefont {Kim}\ \emph {et~al.}(2008)\citenamefont {Kim},
  \citenamefont {Jin}, \citenamefont {Moon}, \citenamefont {Kim}, \citenamefont
  {Park}, \citenamefont {Leem}, \citenamefont {Yu}, \citenamefont {Noh},
  \citenamefont {Kim}, \citenamefont {Oh}, \citenamefont {Park}, \citenamefont
  {Durairaj}, \citenamefont {Cao},\ and\ \citenamefont {Rotenberg}}]{Kim2008}%
  \BibitemOpen
  \bibfield  {author} {\bibinfo {author} {\bibfnamefont {B.~J.}\ \bibnamefont
  {Kim}}, \bibinfo {author} {\bibfnamefont {H.}~\bibnamefont {Jin}}, \bibinfo
  {author} {\bibfnamefont {S.~J.}\ \bibnamefont {Moon}}, \bibinfo {author}
  {\bibfnamefont {J.-Y.}\ \bibnamefont {Kim}}, \bibinfo {author} {\bibfnamefont
  {B.-G.}\ \bibnamefont {Park}}, \bibinfo {author} {\bibfnamefont {C.~S.}\
  \bibnamefont {Leem}}, \bibinfo {author} {\bibfnamefont {J.}~\bibnamefont
  {Yu}}, \bibinfo {author} {\bibfnamefont {T.~W.}\ \bibnamefont {Noh}},
  \bibinfo {author} {\bibfnamefont {C.}~\bibnamefont {Kim}}, \bibinfo {author}
  {\bibfnamefont {S.-J.}\ \bibnamefont {Oh}}, \bibinfo {author} {\bibfnamefont
  {J.-H.}\ \bibnamefont {Park}}, \bibinfo {author} {\bibfnamefont
  {V.}~\bibnamefont {Durairaj}}, \bibinfo {author} {\bibfnamefont
  {G.}~\bibnamefont {Cao}}, \ and\ \bibinfo {author} {\bibfnamefont
  {E.}~\bibnamefont {Rotenberg}},\ }\href {\doibase
  10.1103/PhysRevLett.101.076402} {\bibfield  {journal} {\bibinfo  {journal}
  {Phys. Rev. Lett.}\ }\textbf {\bibinfo {volume} {101}},\ \bibinfo {pages}
  {076402} (\bibinfo {year} {2008})}\BibitemShut {NoStop}%
\bibitem [{\citenamefont {Kim}\ \emph {et~al.}(2009)\citenamefont {Kim},
  \citenamefont {Ohsumi}, \citenamefont {Komesu}, \citenamefont {Sakai},
  \citenamefont {Morita}, \citenamefont {Takagi},\ and\ \citenamefont
  {Arima}}]{Kim2009}%
  \BibitemOpen
  \bibfield  {author} {\bibinfo {author} {\bibfnamefont {B.~J.}\ \bibnamefont
  {Kim}}, \bibinfo {author} {\bibfnamefont {H.}~\bibnamefont {Ohsumi}},
  \bibinfo {author} {\bibfnamefont {T.}~\bibnamefont {Komesu}}, \bibinfo
  {author} {\bibfnamefont {S.}~\bibnamefont {Sakai}}, \bibinfo {author}
  {\bibfnamefont {T.}~\bibnamefont {Morita}}, \bibinfo {author} {\bibfnamefont
  {H.}~\bibnamefont {Takagi}}, \ and\ \bibinfo {author} {\bibfnamefont
  {T.}~\bibnamefont {Arima}},\ }\href {\doibase 10.1126/science.1167106}
  {\bibfield  {journal} {\bibinfo  {journal} {Science}\ }\textbf {\bibinfo
  {volume} {323}},\ \bibinfo {pages} {1329} (\bibinfo {year}
  {2009})}\BibitemShut {NoStop}%
\bibitem [{\citenamefont {Kim}\ \emph {et~al.}(2012{\natexlab{a}})\citenamefont
  {Kim}, \citenamefont {Casa}, \citenamefont {Upton}, \citenamefont {Gog},
  \citenamefont {Kim}, \citenamefont {Mitchell}, \citenamefont {van
  Veenendaal}, \citenamefont {Daghofer}, \citenamefont {van~den Brink},
  \citenamefont {Khaliullin},\ and\ \citenamefont {Kim}}]{Kim2012}%
  \BibitemOpen
  \bibfield  {author} {\bibinfo {author} {\bibfnamefont {J.}~\bibnamefont
  {Kim}}, \bibinfo {author} {\bibfnamefont {D.}~\bibnamefont {Casa}}, \bibinfo
  {author} {\bibfnamefont {M.~H.}\ \bibnamefont {Upton}}, \bibinfo {author}
  {\bibfnamefont {T.}~\bibnamefont {Gog}}, \bibinfo {author} {\bibfnamefont
  {Y.-J.}\ \bibnamefont {Kim}}, \bibinfo {author} {\bibfnamefont {J.~F.}\
  \bibnamefont {Mitchell}}, \bibinfo {author} {\bibfnamefont {M.}~\bibnamefont
  {van Veenendaal}}, \bibinfo {author} {\bibfnamefont {M.}~\bibnamefont
  {Daghofer}}, \bibinfo {author} {\bibfnamefont {J.}~\bibnamefont {van~den
  Brink}}, \bibinfo {author} {\bibfnamefont {G.}~\bibnamefont {Khaliullin}}, \
  and\ \bibinfo {author} {\bibfnamefont {B.~J.}\ \bibnamefont {Kim}},\ }\href
  {\doibase 10.1103/PhysRevLett.108.177003} {\bibfield  {journal} {\bibinfo
  {journal} {Phys. Rev. Lett.}\ }\textbf {\bibinfo {volume} {108}},\ \bibinfo
  {pages} {177003} (\bibinfo {year} {2012}{\natexlab{a}})}\BibitemShut
  {NoStop}%
\bibitem [{\citenamefont {Fujiyama}\ \emph {et~al.}(2012)\citenamefont
  {Fujiyama}, \citenamefont {Ohsumi}, \citenamefont {Komesu}, \citenamefont
  {Matsuno}, \citenamefont {Kim}, \citenamefont {Takata}, \citenamefont
  {Arima},\ and\ \citenamefont {Takagi}}]{Fujiyama}%
  \BibitemOpen
  \bibfield  {author} {\bibinfo {author} {\bibfnamefont {S.}~\bibnamefont
  {Fujiyama}}, \bibinfo {author} {\bibfnamefont {H.}~\bibnamefont {Ohsumi}},
  \bibinfo {author} {\bibfnamefont {T.}~\bibnamefont {Komesu}}, \bibinfo
  {author} {\bibfnamefont {J.}~\bibnamefont {Matsuno}}, \bibinfo {author}
  {\bibfnamefont {B.~J.}\ \bibnamefont {Kim}}, \bibinfo {author} {\bibfnamefont
  {M.}~\bibnamefont {Takata}}, \bibinfo {author} {\bibfnamefont
  {T.}~\bibnamefont {Arima}}, \ and\ \bibinfo {author} {\bibfnamefont
  {H.}~\bibnamefont {Takagi}},\ }\href {\doibase
  10.1103/PhysRevLett.108.247212} {\bibfield  {journal} {\bibinfo  {journal}
  {Phys. Rev. Lett.}\ }\textbf {\bibinfo {volume} {108}},\ \bibinfo {pages}
  {247212} (\bibinfo {year} {2012})}\BibitemShut {NoStop}%
\bibitem [{\citenamefont {Bertinshaw}\ \emph {et~al.}(ress)\citenamefont
  {Bertinshaw}, \citenamefont {Kim}, \citenamefont {Khaliullin},\ and\
  \citenamefont {Kim}}]{Bertinshaw2018}%
  \BibitemOpen
  \bibfield  {author} {\bibinfo {author} {\bibfnamefont {J.}~\bibnamefont
  {Bertinshaw}}, \bibinfo {author} {\bibfnamefont {Y.~K.}\ \bibnamefont {Kim}},
  \bibinfo {author} {\bibfnamefont {G.}~\bibnamefont {Khaliullin}}, \ and\
  \bibinfo {author} {\bibfnamefont {B.~J.}\ \bibnamefont {Kim}},\ }\href@noop
  {} {\bibfield  {journal} {\bibinfo  {journal} {Annu. Rev. Condens. Matter
  Phys.}\ } (\bibinfo {year} {in press})}\BibitemShut {NoStop}%
\bibitem [{\citenamefont {Khaliullin}(2013)}]{Khaliullin2013}%
  \BibitemOpen
  \bibfield  {author} {\bibinfo {author} {\bibfnamefont {G.}~\bibnamefont
  {Khaliullin}},\ }\href {\doibase 10.1103/PhysRevLett.111.197201} {\bibfield
  {journal} {\bibinfo  {journal} {Phys. Rev. Lett.}\ }\textbf {\bibinfo
  {volume} {111}},\ \bibinfo {pages} {197201} (\bibinfo {year}
  {2013})}\BibitemShut {NoStop}%
\bibitem [{\citenamefont {Jain}\ \emph {et~al.}(2017)\citenamefont {Jain},
  \citenamefont {Krautloher}, \citenamefont {Porras}, \citenamefont {Ryu},
  \citenamefont {Chen}, \citenamefont {Abernathy}, \citenamefont {Park},
  \citenamefont {Ivanov}, \citenamefont {Chaloupka}, \citenamefont
  {Khaliullin}, \citenamefont {Keimer},\ and\ \citenamefont {Kim}}]{Jain2017}%
  \BibitemOpen
  \bibfield  {author} {\bibinfo {author} {\bibfnamefont {A.}~\bibnamefont
  {Jain}}, \bibinfo {author} {\bibfnamefont {M.}~\bibnamefont {Krautloher}},
  \bibinfo {author} {\bibfnamefont {J.}~\bibnamefont {Porras}}, \bibinfo
  {author} {\bibfnamefont {G.~H.}\ \bibnamefont {Ryu}}, \bibinfo {author}
  {\bibfnamefont {D.~P.}\ \bibnamefont {Chen}}, \bibinfo {author}
  {\bibfnamefont {D.~L.}\ \bibnamefont {Abernathy}}, \bibinfo {author}
  {\bibfnamefont {J.~T.}\ \bibnamefont {Park}}, \bibinfo {author}
  {\bibfnamefont {A.}~\bibnamefont {Ivanov}}, \bibinfo {author} {\bibfnamefont
  {J.}~\bibnamefont {Chaloupka}}, \bibinfo {author} {\bibfnamefont
  {G.}~\bibnamefont {Khaliullin}}, \bibinfo {author} {\bibfnamefont
  {B.}~\bibnamefont {Keimer}}, \ and\ \bibinfo {author} {\bibfnamefont {B.~J.}\
  \bibnamefont {Kim}},\ }\href {http://dx.doi.org/10.1038/nphys4077} {\bibfield
   {journal} {\bibinfo  {journal} {Nat. Phys.}\ }\textbf {\bibinfo {volume}
  {13}},\ \bibinfo {pages} {633} (\bibinfo {year} {2017})}\BibitemShut
  {NoStop}%
\bibitem [{not()}]{note1}%
  \BibitemOpen
  \href@noop {} {}\bibinfo {note} {See recent works
  \cite{MorettiSala2018,Said,jkim2018} and references therein}\BibitemShut
  {NoStop}%
\bibitem [{\citenamefont {Gretarsson}\ \emph {et~al.}(2016)\citenamefont
  {Gretarsson}, \citenamefont {Sung}, \citenamefont {Porras}, \citenamefont
  {Bertinshaw}, \citenamefont {Dietl}, \citenamefont {Bruin}, \citenamefont
  {Bangura}, \citenamefont {Kim}, \citenamefont {Dinnebier}, \citenamefont
  {Kim}, \citenamefont {Al-Zein}, \citenamefont {Moretti~Sala}, \citenamefont
  {Krisch}, \citenamefont {Le~Tacon}, \citenamefont {Keimer},\ and\
  \citenamefont {Kim}}]{GretarssonRIXS}%
  \BibitemOpen
  \bibfield  {author} {\bibinfo {author} {\bibfnamefont {H.}~\bibnamefont
  {Gretarsson}}, \bibinfo {author} {\bibfnamefont {N.~H.}\ \bibnamefont
  {Sung}}, \bibinfo {author} {\bibfnamefont {J.}~\bibnamefont {Porras}},
  \bibinfo {author} {\bibfnamefont {J.}~\bibnamefont {Bertinshaw}}, \bibinfo
  {author} {\bibfnamefont {C.}~\bibnamefont {Dietl}}, \bibinfo {author}
  {\bibfnamefont {J.~A.~N.}\ \bibnamefont {Bruin}}, \bibinfo {author}
  {\bibfnamefont {A.~F.}\ \bibnamefont {Bangura}}, \bibinfo {author}
  {\bibfnamefont {Y.~K.}\ \bibnamefont {Kim}}, \bibinfo {author} {\bibfnamefont
  {R.}~\bibnamefont {Dinnebier}}, \bibinfo {author} {\bibfnamefont
  {J.}~\bibnamefont {Kim}}, \bibinfo {author} {\bibfnamefont {A.}~\bibnamefont
  {Al-Zein}}, \bibinfo {author} {\bibfnamefont {M.}~\bibnamefont
  {Moretti~Sala}}, \bibinfo {author} {\bibfnamefont {M.}~\bibnamefont
  {Krisch}}, \bibinfo {author} {\bibfnamefont {M.}~\bibnamefont {Le~Tacon}},
  \bibinfo {author} {\bibfnamefont {B.}~\bibnamefont {Keimer}}, \ and\ \bibinfo
  {author} {\bibfnamefont {B.~J.}\ \bibnamefont {Kim}},\ }\href {\doibase
  10.1103/PhysRevLett.117.107001} {\bibfield  {journal} {\bibinfo  {journal}
  {Phys. Rev. Lett.}\ }\textbf {\bibinfo {volume} {117}},\ \bibinfo {pages}
  {107001} (\bibinfo {year} {2016})}\BibitemShut {NoStop}%
\bibitem [{\citenamefont {Pincini}\ \emph {et~al.}(2017)\citenamefont
  {Pincini}, \citenamefont {Vale}, \citenamefont {Donnerer}, \citenamefont
  {de~la Torre}, \citenamefont {Hunter}, \citenamefont {Perry}, \citenamefont
  {Moretti~Sala}, \citenamefont {Baumberger},\ and\ \citenamefont
  {McMorrow}}]{Pincini}%
  \BibitemOpen
  \bibfield  {author} {\bibinfo {author} {\bibfnamefont {D.}~\bibnamefont
  {Pincini}}, \bibinfo {author} {\bibfnamefont {J.~G.}\ \bibnamefont {Vale}},
  \bibinfo {author} {\bibfnamefont {C.}~\bibnamefont {Donnerer}}, \bibinfo
  {author} {\bibfnamefont {A.}~\bibnamefont {de~la Torre}}, \bibinfo {author}
  {\bibfnamefont {E.~C.}\ \bibnamefont {Hunter}}, \bibinfo {author}
  {\bibfnamefont {R.}~\bibnamefont {Perry}}, \bibinfo {author} {\bibfnamefont
  {M.}~\bibnamefont {Moretti~Sala}}, \bibinfo {author} {\bibfnamefont
  {F.}~\bibnamefont {Baumberger}}, \ and\ \bibinfo {author} {\bibfnamefont
  {D.~F.}\ \bibnamefont {McMorrow}},\ }\href {\doibase
  10.1103/PhysRevB.96.075162} {\bibfield  {journal} {\bibinfo  {journal} {Phys.
  Rev. B}\ }\textbf {\bibinfo {volume} {96}},\ \bibinfo {pages} {075162}
  (\bibinfo {year} {2017})}\BibitemShut {NoStop}%
\bibitem [{\citenamefont {Kim}\ \emph {et~al.}(2012{\natexlab{b}})\citenamefont
  {Kim}, \citenamefont {Said}, \citenamefont {Casa}, \citenamefont {Upton},
  \citenamefont {Gog}, \citenamefont {Daghofer}, \citenamefont {Jackeli},
  \citenamefont {van~den Brink}, \citenamefont {Khaliullin},\ and\
  \citenamefont {Kim}}]{PhysRevLett.109.157402}%
  \BibitemOpen
  \bibfield  {author} {\bibinfo {author} {\bibfnamefont {J.}~\bibnamefont
  {Kim}}, \bibinfo {author} {\bibfnamefont {A.~H.}\ \bibnamefont {Said}},
  \bibinfo {author} {\bibfnamefont {D.}~\bibnamefont {Casa}}, \bibinfo {author}
  {\bibfnamefont {M.~H.}\ \bibnamefont {Upton}}, \bibinfo {author}
  {\bibfnamefont {T.}~\bibnamefont {Gog}}, \bibinfo {author} {\bibfnamefont
  {M.}~\bibnamefont {Daghofer}}, \bibinfo {author} {\bibfnamefont
  {G.}~\bibnamefont {Jackeli}}, \bibinfo {author} {\bibfnamefont
  {J.}~\bibnamefont {van~den Brink}}, \bibinfo {author} {\bibfnamefont
  {G.}~\bibnamefont {Khaliullin}}, \ and\ \bibinfo {author} {\bibfnamefont
  {B.~J.}\ \bibnamefont {Kim}},\ }\href {\doibase
  10.1103/PhysRevLett.109.157402} {\bibfield  {journal} {\bibinfo  {journal}
  {Phys. Rev. Lett.}\ }\textbf {\bibinfo {volume} {109}},\ \bibinfo {pages}
  {157402} (\bibinfo {year} {2012}{\natexlab{b}})}\BibitemShut {NoStop}%
\bibitem [{\citenamefont {Wan}\ \emph {et~al.}(2011)\citenamefont {Wan},
  \citenamefont {Turner}, \citenamefont {Vishwanath},\ and\ \citenamefont
  {Savrasov}}]{Wan2011}%
  \BibitemOpen
  \bibfield  {author} {\bibinfo {author} {\bibfnamefont {X.}~\bibnamefont
  {Wan}}, \bibinfo {author} {\bibfnamefont {A.~M.}\ \bibnamefont {Turner}},
  \bibinfo {author} {\bibfnamefont {A.}~\bibnamefont {Vishwanath}}, \ and\
  \bibinfo {author} {\bibfnamefont {S.~Y.}\ \bibnamefont {Savrasov}},\ }\href
  {\doibase 10.1103/PhysRevB.83.205101} {\bibfield  {journal} {\bibinfo
  {journal} {Phys. Rev. B}\ }\textbf {\bibinfo {volume} {83}},\ \bibinfo
  {pages} {205101} (\bibinfo {year} {2011})}\BibitemShut {NoStop}%
\bibitem [{\citenamefont {Dean}\ \emph {et~al.}(2016)\citenamefont {Dean} \emph
  {et~al.}}]{Dean2016}%
  \BibitemOpen
  \bibfield  {author} {\bibinfo {author} {\bibfnamefont {M.~P.~M.}\
  \bibnamefont {Dean}} \emph {et~al.},\ }\href
  {http://dx.doi.org/10.1038/nmat4641} {\bibfield  {journal} {\bibinfo
  {journal} {Nature Materials}\ }\textbf {\bibinfo {volume} {15}},\ \bibinfo
  {pages} {601} (\bibinfo {year} {2016})}\BibitemShut {NoStop}%
\bibitem [{\citenamefont {Kim}\ \emph {et~al.}(2014)\citenamefont {Kim},
  \citenamefont {Krupin}, \citenamefont {Denlinger}, \citenamefont {Bostwick},
  \citenamefont {Rotenberg}, \citenamefont {Zhao}, \citenamefont {Mitchell},
  \citenamefont {Allen},\ and\ \citenamefont {Kim}}]{Kim2014}%
  \BibitemOpen
  \bibfield  {author} {\bibinfo {author} {\bibfnamefont {Y.~K.}\ \bibnamefont
  {Kim}}, \bibinfo {author} {\bibfnamefont {O.}~\bibnamefont {Krupin}},
  \bibinfo {author} {\bibfnamefont {J.~D.}\ \bibnamefont {Denlinger}}, \bibinfo
  {author} {\bibfnamefont {A.}~\bibnamefont {Bostwick}}, \bibinfo {author}
  {\bibfnamefont {E.}~\bibnamefont {Rotenberg}}, \bibinfo {author}
  {\bibfnamefont {Q.}~\bibnamefont {Zhao}}, \bibinfo {author} {\bibfnamefont
  {J.~F.}\ \bibnamefont {Mitchell}}, \bibinfo {author} {\bibfnamefont {J.~W.}\
  \bibnamefont {Allen}}, \ and\ \bibinfo {author} {\bibfnamefont {B.~J.}\
  \bibnamefont {Kim}},\ }\href {\doibase 10.1126/science.1251151} {\bibfield
  {journal} {\bibinfo  {journal} {Science}\ }\textbf {\bibinfo {volume}
  {345}},\ \bibinfo {pages} {187} (\bibinfo {year} {2014})}\BibitemShut
  {NoStop}%
\bibitem [{\citenamefont {Kim}\ \emph {et~al.}(2015)\citenamefont {Kim},
  \citenamefont {Sung}, \citenamefont {Denlinger},\ and\ \citenamefont
  {Kim}}]{Kim2015}%
  \BibitemOpen
  \bibfield  {author} {\bibinfo {author} {\bibfnamefont {Y.~K.}\ \bibnamefont
  {Kim}}, \bibinfo {author} {\bibfnamefont {N.~H.}\ \bibnamefont {Sung}},
  \bibinfo {author} {\bibfnamefont {J.~D.}\ \bibnamefont {Denlinger}}, \ and\
  \bibinfo {author} {\bibfnamefont {B.~J.}\ \bibnamefont {Kim}},\ }\href
  {http://dx.doi.org/10.1038/nphys3503} {\bibfield  {journal} {\bibinfo
  {journal} {Nat. Phys.}\ }\textbf {\bibinfo {volume} {12}},\ \bibinfo {pages}
  {37} (\bibinfo {year} {2015})}\BibitemShut {NoStop}%
\bibitem [{\citenamefont {Yan}\ \emph {et~al.}(2015)\citenamefont {Yan},
  \citenamefont {Ren}, \citenamefont {Xu}, \citenamefont {Xie}, \citenamefont
  {Tao}, \citenamefont {Choi}, \citenamefont {Lee}, \citenamefont {Choi},
  \citenamefont {Zhang},\ and\ \citenamefont {Feng}}]{Yan}%
  \BibitemOpen
  \bibfield  {author} {\bibinfo {author} {\bibfnamefont {Y.~J.}\ \bibnamefont
  {Yan}}, \bibinfo {author} {\bibfnamefont {M.~Q.}\ \bibnamefont {Ren}},
  \bibinfo {author} {\bibfnamefont {H.~C.}\ \bibnamefont {Xu}}, \bibinfo
  {author} {\bibfnamefont {B.~P.}\ \bibnamefont {Xie}}, \bibinfo {author}
  {\bibfnamefont {R.}~\bibnamefont {Tao}}, \bibinfo {author} {\bibfnamefont
  {H.~Y.}\ \bibnamefont {Choi}}, \bibinfo {author} {\bibfnamefont
  {N.}~\bibnamefont {Lee}}, \bibinfo {author} {\bibfnamefont {Y.~J.}\
  \bibnamefont {Choi}}, \bibinfo {author} {\bibfnamefont {T.}~\bibnamefont
  {Zhang}}, \ and\ \bibinfo {author} {\bibfnamefont {D.~L.}\ \bibnamefont
  {Feng}},\ }\href {\doibase 10.1103/PhysRevX.5.041018} {\bibfield  {journal}
  {\bibinfo  {journal} {Phys. Rev. X}\ }\textbf {\bibinfo {volume} {5}},\
  \bibinfo {pages} {041018} (\bibinfo {year} {2015})}\BibitemShut {NoStop}%
\bibitem [{\citenamefont {Crawford}\ \emph {et~al.}(1994)\citenamefont
  {Crawford}, \citenamefont {Subramanian}, \citenamefont {Harlow},
  \citenamefont {Fernandez-Baca}, \citenamefont {Wang},\ and\ \citenamefont
  {Johnston}}]{Crawford}%
  \BibitemOpen
  \bibfield  {author} {\bibinfo {author} {\bibfnamefont {M.~K.}\ \bibnamefont
  {Crawford}}, \bibinfo {author} {\bibfnamefont {M.~A.}\ \bibnamefont
  {Subramanian}}, \bibinfo {author} {\bibfnamefont {R.~L.}\ \bibnamefont
  {Harlow}}, \bibinfo {author} {\bibfnamefont {J.~A.}\ \bibnamefont
  {Fernandez-Baca}}, \bibinfo {author} {\bibfnamefont {Z.~R.}\ \bibnamefont
  {Wang}}, \ and\ \bibinfo {author} {\bibfnamefont {D.~C.}\ \bibnamefont
  {Johnston}},\ }\href {\doibase 10.1103/PhysRevB.49.9198} {\bibfield
  {journal} {\bibinfo  {journal} {Phys. Rev. B}\ }\textbf {\bibinfo {volume}
  {49}},\ \bibinfo {pages} {9198} (\bibinfo {year} {1994})}\BibitemShut
  {NoStop}%
\bibitem [{\citenamefont {Dhital}\ \emph {et~al.}(2013)\citenamefont {Dhital},
  \citenamefont {Hogan}, \citenamefont {Yamani}, \citenamefont {de~la Cruz},
  \citenamefont {Chen}, \citenamefont {Khadka}, \citenamefont {Ren},\ and\
  \citenamefont {Wilson}}]{Dhital}%
  \BibitemOpen
  \bibfield  {author} {\bibinfo {author} {\bibfnamefont {C.}~\bibnamefont
  {Dhital}}, \bibinfo {author} {\bibfnamefont {T.}~\bibnamefont {Hogan}},
  \bibinfo {author} {\bibfnamefont {Z.}~\bibnamefont {Yamani}}, \bibinfo
  {author} {\bibfnamefont {C.}~\bibnamefont {de~la Cruz}}, \bibinfo {author}
  {\bibfnamefont {X.}~\bibnamefont {Chen}}, \bibinfo {author} {\bibfnamefont
  {S.}~\bibnamefont {Khadka}}, \bibinfo {author} {\bibfnamefont
  {Z.}~\bibnamefont {Ren}}, \ and\ \bibinfo {author} {\bibfnamefont {S.~D.}\
  \bibnamefont {Wilson}},\ }\href {\doibase 10.1103/PhysRevB.87.144405}
  {\bibfield  {journal} {\bibinfo  {journal} {Phys. Rev. B}\ }\textbf {\bibinfo
  {volume} {87}},\ \bibinfo {pages} {144405} (\bibinfo {year}
  {2013})}\BibitemShut {NoStop}%
\bibitem [{\citenamefont {Ye}\ \emph {et~al.}(2013)\citenamefont {Ye},
  \citenamefont {Chi}, \citenamefont {Chakoumakos}, \citenamefont
  {Fernandez-Baca}, \citenamefont {Qi},\ and\ \citenamefont {Cao}}]{Ye}%
  \BibitemOpen
  \bibfield  {author} {\bibinfo {author} {\bibfnamefont {F.}~\bibnamefont
  {Ye}}, \bibinfo {author} {\bibfnamefont {S.}~\bibnamefont {Chi}}, \bibinfo
  {author} {\bibfnamefont {B.~C.}\ \bibnamefont {Chakoumakos}}, \bibinfo
  {author} {\bibfnamefont {J.~A.}\ \bibnamefont {Fernandez-Baca}}, \bibinfo
  {author} {\bibfnamefont {T.}~\bibnamefont {Qi}}, \ and\ \bibinfo {author}
  {\bibfnamefont {G.}~\bibnamefont {Cao}},\ }\href {\doibase
  10.1103/PhysRevB.87.140406} {\bibfield  {journal} {\bibinfo  {journal} {Phys.
  Rev. B}\ }\textbf {\bibinfo {volume} {87}},\ \bibinfo {pages} {140406}
  (\bibinfo {year} {2013})}\BibitemShut {NoStop}%
\bibitem [{\citenamefont {Vale}\ \emph {et~al.}(2015)\citenamefont {Vale},
  \citenamefont {Boseggia}, \citenamefont {Walker}, \citenamefont {Springell},
  \citenamefont {Feng}, \citenamefont {Hunter}, \citenamefont {Perry},
  \citenamefont {Prabhakaran}, \citenamefont {Boothroyd}, \citenamefont
  {Collins}, \citenamefont {R\o{}nnow},\ and\ \citenamefont {McMorrow}}]{Vale}%
  \BibitemOpen
  \bibfield  {author} {\bibinfo {author} {\bibfnamefont {J.~G.}\ \bibnamefont
  {Vale}}, \bibinfo {author} {\bibfnamefont {S.}~\bibnamefont {Boseggia}},
  \bibinfo {author} {\bibfnamefont {H.~C.}\ \bibnamefont {Walker}}, \bibinfo
  {author} {\bibfnamefont {R.}~\bibnamefont {Springell}}, \bibinfo {author}
  {\bibfnamefont {Z.}~\bibnamefont {Feng}}, \bibinfo {author} {\bibfnamefont
  {E.~C.}\ \bibnamefont {Hunter}}, \bibinfo {author} {\bibfnamefont {R.~S.}\
  \bibnamefont {Perry}}, \bibinfo {author} {\bibfnamefont {D.}~\bibnamefont
  {Prabhakaran}}, \bibinfo {author} {\bibfnamefont {A.~T.}\ \bibnamefont
  {Boothroyd}}, \bibinfo {author} {\bibfnamefont {S.~P.}\ \bibnamefont
  {Collins}}, \bibinfo {author} {\bibfnamefont {H.~M.}\ \bibnamefont
  {R\o{}nnow}}, \ and\ \bibinfo {author} {\bibfnamefont {D.~F.}\ \bibnamefont
  {McMorrow}},\ }\href {\doibase 10.1103/PhysRevB.92.020406} {\bibfield
  {journal} {\bibinfo  {journal} {Phys. Rev. B}\ }\textbf {\bibinfo {volume}
  {92}},\ \bibinfo {pages} {020406} (\bibinfo {year} {2015})}\BibitemShut
  {NoStop}%
\bibitem [{\citenamefont {Bahr}\ \emph {et~al.}(2014)\citenamefont {Bahr},
  \citenamefont {Alfonsov}, \citenamefont {Jackeli}, \citenamefont
  {Khaliullin}, \citenamefont {Matsumoto}, \citenamefont {Takayama},
  \citenamefont {Takagi}, \citenamefont {B\"uchner},\ and\ \citenamefont
  {Kataev}}]{Bahr}%
  \BibitemOpen
  \bibfield  {author} {\bibinfo {author} {\bibfnamefont {S.}~\bibnamefont
  {Bahr}}, \bibinfo {author} {\bibfnamefont {A.}~\bibnamefont {Alfonsov}},
  \bibinfo {author} {\bibfnamefont {G.}~\bibnamefont {Jackeli}}, \bibinfo
  {author} {\bibfnamefont {G.}~\bibnamefont {Khaliullin}}, \bibinfo {author}
  {\bibfnamefont {A.}~\bibnamefont {Matsumoto}}, \bibinfo {author}
  {\bibfnamefont {T.}~\bibnamefont {Takayama}}, \bibinfo {author}
  {\bibfnamefont {H.}~\bibnamefont {Takagi}}, \bibinfo {author} {\bibfnamefont
  {B.}~\bibnamefont {B\"uchner}}, \ and\ \bibinfo {author} {\bibfnamefont
  {V.}~\bibnamefont {Kataev}},\ }\href {\doibase 10.1103/PhysRevB.89.180401}
  {\bibfield  {journal} {\bibinfo  {journal} {Phys. Rev. B}\ }\textbf {\bibinfo
  {volume} {89}},\ \bibinfo {pages} {180401} (\bibinfo {year}
  {2014})}\BibitemShut {NoStop}%
\bibitem [{\citenamefont {Gim}\ \emph {et~al.}(2016)\citenamefont {Gim},
  \citenamefont {Sethi}, \citenamefont {Zhao}, \citenamefont {Mitchell},
  \citenamefont {Cao},\ and\ \citenamefont {Cooper}}]{Gim}%
  \BibitemOpen
  \bibfield  {author} {\bibinfo {author} {\bibfnamefont {Y.}~\bibnamefont
  {Gim}}, \bibinfo {author} {\bibfnamefont {A.}~\bibnamefont {Sethi}}, \bibinfo
  {author} {\bibfnamefont {Q.}~\bibnamefont {Zhao}}, \bibinfo {author}
  {\bibfnamefont {J.~F.}\ \bibnamefont {Mitchell}}, \bibinfo {author}
  {\bibfnamefont {G.}~\bibnamefont {Cao}}, \ and\ \bibinfo {author}
  {\bibfnamefont {S.~L.}\ \bibnamefont {Cooper}},\ }\href {\doibase
  10.1103/PhysRevB.93.024405} {\bibfield  {journal} {\bibinfo  {journal} {Phys.
  Rev. B}\ }\textbf {\bibinfo {volume} {93}},\ \bibinfo {pages} {024405}
  (\bibinfo {year} {2016})}\BibitemShut {NoStop}%
\bibitem [{\citenamefont {Gretarsson}\ \emph {et~al.}(2017)\citenamefont
  {Gretarsson}, \citenamefont {Sauceda}, \citenamefont {Sung}, \citenamefont
  {H\"oppner}, \citenamefont {Minola}, \citenamefont {Kim}, \citenamefont
  {Keimer},\ and\ \citenamefont {Le~Tacon}}]{Gretarsson2017}%
  \BibitemOpen
  \bibfield  {author} {\bibinfo {author} {\bibfnamefont {H.}~\bibnamefont
  {Gretarsson}}, \bibinfo {author} {\bibfnamefont {J.}~\bibnamefont {Sauceda}},
  \bibinfo {author} {\bibfnamefont {N.~H.}\ \bibnamefont {Sung}}, \bibinfo
  {author} {\bibfnamefont {M.}~\bibnamefont {H\"oppner}}, \bibinfo {author}
  {\bibfnamefont {M.}~\bibnamefont {Minola}}, \bibinfo {author} {\bibfnamefont
  {B.~J.}\ \bibnamefont {Kim}}, \bibinfo {author} {\bibfnamefont
  {B.}~\bibnamefont {Keimer}}, \ and\ \bibinfo {author} {\bibfnamefont
  {M.}~\bibnamefont {Le~Tacon}},\ }\href {\doibase 10.1103/PhysRevB.96.115138}
  {\bibfield  {journal} {\bibinfo  {journal} {Phys. Rev. B}\ }\textbf {\bibinfo
  {volume} {96}},\ \bibinfo {pages} {115138} (\bibinfo {year}
  {2017})}\BibitemShut {NoStop}%
\bibitem [{\citenamefont {Di~Matteo}\ and\ \citenamefont
  {Norman}(2016)}]{DiMatteo}%
  \BibitemOpen
  \bibfield  {author} {\bibinfo {author} {\bibfnamefont {S.}~\bibnamefont
  {Di~Matteo}}\ and\ \bibinfo {author} {\bibfnamefont {M.~R.}\ \bibnamefont
  {Norman}},\ }\href {\doibase 10.1103/PhysRevB.94.075148} {\bibfield
  {journal} {\bibinfo  {journal} {Phys. Rev. B}\ }\textbf {\bibinfo {volume}
  {94}},\ \bibinfo {pages} {075148} (\bibinfo {year} {2016})}\BibitemShut
  {NoStop}%
\bibitem [{\citenamefont {Torchinsky}\ \emph {et~al.}(2015)\citenamefont
  {Torchinsky}, \citenamefont {Chu}, \citenamefont {Zhao}, \citenamefont
  {Perkins}, \citenamefont {Sizyuk}, \citenamefont {Qi}, \citenamefont {Cao},\
  and\ \citenamefont {Hsieh}}]{Torchinsky}%
  \BibitemOpen
  \bibfield  {author} {\bibinfo {author} {\bibfnamefont {D.~H.}\ \bibnamefont
  {Torchinsky}}, \bibinfo {author} {\bibfnamefont {H.}~\bibnamefont {Chu}},
  \bibinfo {author} {\bibfnamefont {L.}~\bibnamefont {Zhao}}, \bibinfo {author}
  {\bibfnamefont {N.~B.}\ \bibnamefont {Perkins}}, \bibinfo {author}
  {\bibfnamefont {Y.}~\bibnamefont {Sizyuk}}, \bibinfo {author} {\bibfnamefont
  {T.}~\bibnamefont {Qi}}, \bibinfo {author} {\bibfnamefont {G.}~\bibnamefont
  {Cao}}, \ and\ \bibinfo {author} {\bibfnamefont {D.}~\bibnamefont {Hsieh}},\
  }\href {\doibase 10.1103/PhysRevLett.114.096404} {\bibfield  {journal}
  {\bibinfo  {journal} {Phys. Rev. Lett.}\ }\textbf {\bibinfo {volume} {114}},\
  \bibinfo {pages} {096404} (\bibinfo {year} {2015})}\BibitemShut {NoStop}%
\bibitem [{\citenamefont {Zhao}\ \emph {et~al.}(2015)\citenamefont {Zhao},
  \citenamefont {Torchinsky}, \citenamefont {Chu}, \citenamefont {Ivanov},
  \citenamefont {Lifshitz}, \citenamefont {Flint}, \citenamefont {Qi},
  \citenamefont {Cao},\ and\ \citenamefont {Hsieh}}]{Zhao}%
  \BibitemOpen
  \bibfield  {author} {\bibinfo {author} {\bibfnamefont {L.}~\bibnamefont
  {Zhao}}, \bibinfo {author} {\bibfnamefont {D.~H.}\ \bibnamefont
  {Torchinsky}}, \bibinfo {author} {\bibfnamefont {H.}~\bibnamefont {Chu}},
  \bibinfo {author} {\bibfnamefont {V.}~\bibnamefont {Ivanov}}, \bibinfo
  {author} {\bibfnamefont {R.}~\bibnamefont {Lifshitz}}, \bibinfo {author}
  {\bibfnamefont {R.}~\bibnamefont {Flint}}, \bibinfo {author} {\bibfnamefont
  {T.}~\bibnamefont {Qi}}, \bibinfo {author} {\bibfnamefont {G.}~\bibnamefont
  {Cao}}, \ and\ \bibinfo {author} {\bibfnamefont {D.}~\bibnamefont {Hsieh}},\
  }\href {http://dx.doi.org/10.1038/nphys3517} {\bibfield  {journal} {\bibinfo
  {journal} {Nat. Phys.}\ }\textbf {\bibinfo {volume} {12}},\ \bibinfo {pages}
  {32} (\bibinfo {year} {2015})}\BibitemShut {NoStop}%
\bibitem [{\citenamefont {Millis}\ \emph {et~al.}(1995)\citenamefont {Millis},
  \citenamefont {Littlewood},\ and\ \citenamefont {Shraiman}}]{Millis}%
  \BibitemOpen
  \bibfield  {author} {\bibinfo {author} {\bibfnamefont {A.~J.}\ \bibnamefont
  {Millis}}, \bibinfo {author} {\bibfnamefont {P.~B.}\ \bibnamefont
  {Littlewood}}, \ and\ \bibinfo {author} {\bibfnamefont {B.~I.}\ \bibnamefont
  {Shraiman}},\ }\href {\doibase 10.1103/PhysRevLett.74.5144} {\bibfield
  {journal} {\bibinfo  {journal} {Phys. Rev. Lett.}\ }\textbf {\bibinfo
  {volume} {74}},\ \bibinfo {pages} {5144} (\bibinfo {year}
  {1995})}\BibitemShut {NoStop}%
\bibitem [{\citenamefont {Gunnarsson}\ and\ \citenamefont
  {R\"osch}(2008)}]{Gunnarsson2008}%
  \BibitemOpen
  \bibfield  {author} {\bibinfo {author} {\bibfnamefont {O.}~\bibnamefont
  {Gunnarsson}}\ and\ \bibinfo {author} {\bibfnamefont {O.}~\bibnamefont
  {R\"osch}},\ }\href {http://stacks.iop.org/0953-8984/20/i=4/a=043201}
  {\bibfield  {journal} {\bibinfo  {journal} {J. Phys.: Condens. Matter}\
  }\textbf {\bibinfo {volume} {20}},\ \bibinfo {pages} {043201} (\bibinfo
  {year} {2008})}\BibitemShut {NoStop}%
\bibitem [{\citenamefont {Takayama}\ \emph {et~al.}(2016)\citenamefont
  {Takayama}, \citenamefont {Matsumoto}, \citenamefont {Jackeli},\ and\
  \citenamefont {Takagi}}]{Takayama}%
  \BibitemOpen
  \bibfield  {author} {\bibinfo {author} {\bibfnamefont {T.}~\bibnamefont
  {Takayama}}, \bibinfo {author} {\bibfnamefont {A.}~\bibnamefont {Matsumoto}},
  \bibinfo {author} {\bibfnamefont {G.}~\bibnamefont {Jackeli}}, \ and\
  \bibinfo {author} {\bibfnamefont {H.}~\bibnamefont {Takagi}},\ }\href
  {\doibase 10.1103/PhysRevB.94.224420} {\bibfield  {journal} {\bibinfo
  {journal} {Phys. Rev. B}\ }\textbf {\bibinfo {volume} {94}},\ \bibinfo
  {pages} {224420} (\bibinfo {year} {2016})}\BibitemShut {NoStop}%
\bibitem [{\citenamefont {Chernyshev}(2005)}]{Cher05}%
  \BibitemOpen
  \bibfield  {author} {\bibinfo {author} {\bibfnamefont {A.~L.}\ \bibnamefont
  {Chernyshev}},\ }\href {\doibase 10.1103/PhysRevB.72.174414} {\bibfield
  {journal} {\bibinfo  {journal} {Phys. Rev. B}\ }\textbf {\bibinfo {volume}
  {72}},\ \bibinfo {pages} {174414} (\bibinfo {year} {2005})}\BibitemShut
  {NoStop}%
\bibitem [{\citenamefont {Lumsden}\ \emph {et~al.}(2001)\citenamefont
  {Lumsden}, \citenamefont {Sales}, \citenamefont {Mandrus}, \citenamefont
  {Nagler},\ and\ \citenamefont {Thompson}}]{Lums01}%
  \BibitemOpen
  \bibfield  {author} {\bibinfo {author} {\bibfnamefont {M.~D.}\ \bibnamefont
  {Lumsden}}, \bibinfo {author} {\bibfnamefont {B.~C.}\ \bibnamefont {Sales}},
  \bibinfo {author} {\bibfnamefont {D.}~\bibnamefont {Mandrus}}, \bibinfo
  {author} {\bibfnamefont {S.~E.}\ \bibnamefont {Nagler}}, \ and\ \bibinfo
  {author} {\bibfnamefont {J.~R.}\ \bibnamefont {Thompson}},\ }\href {\doibase
  10.1103/PhysRevLett.86.159} {\bibfield  {journal} {\bibinfo  {journal} {Phys.
  Rev. Lett.}\ }\textbf {\bibinfo {volume} {86}},\ \bibinfo {pages} {159}
  (\bibinfo {year} {2001})}\BibitemShut {NoStop}%
\bibitem [{\citenamefont {Wang}\ \emph {et~al.}(2014)\citenamefont {Wang},
  \citenamefont {Seinige}, \citenamefont {Cao}, \citenamefont {Zhou},
  \citenamefont {Goodenough},\ and\ \citenamefont {Tsoi}}]{Wang}%
  \BibitemOpen
  \bibfield  {author} {\bibinfo {author} {\bibfnamefont {C.}~\bibnamefont
  {Wang}}, \bibinfo {author} {\bibfnamefont {H.}~\bibnamefont {Seinige}},
  \bibinfo {author} {\bibfnamefont {G.}~\bibnamefont {Cao}}, \bibinfo {author}
  {\bibfnamefont {J.-S.}\ \bibnamefont {Zhou}}, \bibinfo {author}
  {\bibfnamefont {J.~B.}\ \bibnamefont {Goodenough}}, \ and\ \bibinfo {author}
  {\bibfnamefont {M.}~\bibnamefont {Tsoi}},\ }\href {\doibase
  10.1103/PhysRevX.4.041034} {\bibfield  {journal} {\bibinfo  {journal} {Phys.
  Rev. X}\ }\textbf {\bibinfo {volume} {4}},\ \bibinfo {pages} {041034}
  (\bibinfo {year} {2014})}\BibitemShut {NoStop}%
\bibitem [{\citenamefont {Nauman}\ \emph {et~al.}(2017)\citenamefont {Nauman},
  \citenamefont {Hong}, \citenamefont {Hussain}, \citenamefont {Seo},
  \citenamefont {Park}, \citenamefont {Lee}, \citenamefont {Choi},
  \citenamefont {Kang},\ and\ \citenamefont {Jo}}]{Nauman}%
  \BibitemOpen
  \bibfield  {author} {\bibinfo {author} {\bibfnamefont {M.}~\bibnamefont
  {Nauman}}, \bibinfo {author} {\bibfnamefont {Y.}~\bibnamefont {Hong}},
  \bibinfo {author} {\bibfnamefont {T.}~\bibnamefont {Hussain}}, \bibinfo
  {author} {\bibfnamefont {M.~S.}\ \bibnamefont {Seo}}, \bibinfo {author}
  {\bibfnamefont {S.~Y.}\ \bibnamefont {Park}}, \bibinfo {author}
  {\bibfnamefont {N.}~\bibnamefont {Lee}}, \bibinfo {author} {\bibfnamefont
  {Y.~J.}\ \bibnamefont {Choi}}, \bibinfo {author} {\bibfnamefont
  {W.}~\bibnamefont {Kang}}, \ and\ \bibinfo {author} {\bibfnamefont
  {Y.}~\bibnamefont {Jo}},\ }\href {\doibase 10.1103/PhysRevB.96.155102}
  {\bibfield  {journal} {\bibinfo  {journal} {Phys. Rev. B}\ }\textbf {\bibinfo
  {volume} {96}},\ \bibinfo {pages} {155102} (\bibinfo {year}
  {2017})}\BibitemShut {NoStop}%
\bibitem [{\citenamefont {Fruchter}\ \emph {et~al.}(2016)\citenamefont
  {Fruchter}, \citenamefont {Colson},\ and\ \citenamefont {Brouet}}]{Fruchter}%
  \BibitemOpen
  \bibfield  {author} {\bibinfo {author} {\bibfnamefont {L.}~\bibnamefont
  {Fruchter}}, \bibinfo {author} {\bibfnamefont {D.}~\bibnamefont {Colson}}, \
  and\ \bibinfo {author} {\bibfnamefont {V.}~\bibnamefont {Brouet}},\ }\href
  {http://stacks.iop.org/0953-8984/28/i=12/a=126003} {\bibfield  {journal}
  {\bibinfo  {journal} {J. Phys.: Condens. Matter}\ }\textbf {\bibinfo {volume}
  {28}},\ \bibinfo {pages} {126003} (\bibinfo {year} {2016})}\BibitemShut
  {NoStop}%
\bibitem [{\citenamefont {Yildirim}\ \emph {et~al.}(1995)\citenamefont
  {Yildirim}, \citenamefont {Harris}, \citenamefont {Aharony},\ and\
  \citenamefont {Entin-Wohlman}}]{Yildirim}%
  \BibitemOpen
  \bibfield  {author} {\bibinfo {author} {\bibfnamefont {T.}~\bibnamefont
  {Yildirim}}, \bibinfo {author} {\bibfnamefont {A.~B.}\ \bibnamefont
  {Harris}}, \bibinfo {author} {\bibfnamefont {A.}~\bibnamefont {Aharony}}, \
  and\ \bibinfo {author} {\bibfnamefont {O.}~\bibnamefont {Entin-Wohlman}},\
  }\href {\doibase 10.1103/PhysRevB.52.10239} {\bibfield  {journal} {\bibinfo
  {journal} {Phys. Rev. B}\ }\textbf {\bibinfo {volume} {52}},\ \bibinfo
  {pages} {10239} (\bibinfo {year} {1995})}\BibitemShut {NoStop}%
\bibitem [{\citenamefont {Katukuri}\ \emph {et~al.}(2014)\citenamefont
  {Katukuri}, \citenamefont {Yushankhai}, \citenamefont {Siurakshina},
  \citenamefont {van~den Brink}, \citenamefont {Hozoi},\ and\ \citenamefont
  {Rousochatzakis}}]{Katukuri}%
  \BibitemOpen
  \bibfield  {author} {\bibinfo {author} {\bibfnamefont {V.~M.}\ \bibnamefont
  {Katukuri}}, \bibinfo {author} {\bibfnamefont {V.}~\bibnamefont
  {Yushankhai}}, \bibinfo {author} {\bibfnamefont {L.}~\bibnamefont
  {Siurakshina}}, \bibinfo {author} {\bibfnamefont {J.}~\bibnamefont {van~den
  Brink}}, \bibinfo {author} {\bibfnamefont {L.}~\bibnamefont {Hozoi}}, \ and\
  \bibinfo {author} {\bibfnamefont {I.}~\bibnamefont {Rousochatzakis}},\ }\href
  {\doibase 10.1103/PhysRevX.4.021051} {\bibfield  {journal} {\bibinfo
  {journal} {Phys. Rev. X}\ }\textbf {\bibinfo {volume} {4}},\ \bibinfo {pages}
  {021051} (\bibinfo {year} {2014})}\BibitemShut {NoStop}%
\bibitem [{\citenamefont {Liu}\ and\ \citenamefont {Khaliullin}()}]{Liu}%
  \BibitemOpen
  \bibfield  {author} {\bibinfo {author} {\bibfnamefont {H.}~\bibnamefont
  {Liu}}\ and\ \bibinfo {author} {\bibfnamefont {G.}~\bibnamefont
  {Khaliullin}},\ }\href@noop {} {\bibinfo  {journal} {ArXiv:1808.06919}\
  }\BibitemShut {NoStop}%
\bibitem [{\citenamefont {Toth}\ and\ \citenamefont {Lake}(2015)}]{spinw}%
  \BibitemOpen
\bibfield  {journal} {  }\bibfield  {author} {\bibinfo {author} {\bibfnamefont
  {S.}~\bibnamefont {Toth}}\ and\ \bibinfo {author} {\bibfnamefont
  {B.}~\bibnamefont {Lake}},\ }\href
  {http://stacks.iop.org/0953-8984/27/i=16/a=166002} {\bibfield  {journal}
  {\bibinfo  {journal} {J. Phys.: Condens. Matter}\ }\textbf {\bibinfo {volume}
  {27}},\ \bibinfo {pages} {166002} (\bibinfo {year} {2015})}\BibitemShut
  {NoStop}%
\bibitem [{\citenamefont {Jiang}\ and\ \citenamefont {Wiese}(2011)}]{Jiang}%
  \BibitemOpen
  \bibfield  {author} {\bibinfo {author} {\bibfnamefont {F.-J.}\ \bibnamefont
  {Jiang}}\ and\ \bibinfo {author} {\bibfnamefont {U.-J.}\ \bibnamefont
  {Wiese}},\ }\href {\doibase 10.1103/PhysRevB.83.155120} {\bibfield  {journal}
  {\bibinfo  {journal} {Phys. Rev. B}\ }\textbf {\bibinfo {volume} {83}},\
  \bibinfo {pages} {155120} (\bibinfo {year} {2011})}\BibitemShut {NoStop}%
\bibitem [{\citenamefont {Said}\ \emph {et~al.}(2018)\citenamefont {Said},
  \citenamefont {Gog}, \citenamefont {Wieczorek}, \citenamefont {Huang},
  \citenamefont {Casa}, \citenamefont {Kasman}, \citenamefont {Divan},\ and\
  \citenamefont {Kim}}]{Said}%
  \BibitemOpen
  \bibfield  {author} {\bibinfo {author} {\bibfnamefont {A.~H.}\ \bibnamefont
  {Said}}, \bibinfo {author} {\bibfnamefont {T.}~\bibnamefont {Gog}}, \bibinfo
  {author} {\bibfnamefont {M.}~\bibnamefont {Wieczorek}}, \bibinfo {author}
  {\bibfnamefont {X.}~\bibnamefont {Huang}}, \bibinfo {author} {\bibfnamefont
  {D.}~\bibnamefont {Casa}}, \bibinfo {author} {\bibfnamefont {E.}~\bibnamefont
  {Kasman}}, \bibinfo {author} {\bibfnamefont {R.}~\bibnamefont {Divan}}, \
  and\ \bibinfo {author} {\bibfnamefont {J.~H.}\ \bibnamefont {Kim}},\ }\href
  {\doibase 10.1107/S1600577517018185} {\bibfield  {journal} {\bibinfo
  {journal} {J. Synchrotron Rad}\ }\textbf {\bibinfo {volume} {25}},\ \bibinfo
  {pages} {373} (\bibinfo {year} {2018})}\BibitemShut {NoStop}%
\bibitem [{\citenamefont {Kim}\ \emph {et~al.}(2018)\citenamefont {Kim},
  \citenamefont {Casa}, \citenamefont {Said}, \citenamefont {Krakora},
  \citenamefont {Kim}, \citenamefont {Kasman}, \citenamefont {Huang},\ and\
  \citenamefont {Gog}}]{jkim2018}%
  \BibitemOpen
  \bibfield  {author} {\bibinfo {author} {\bibfnamefont {J.}~\bibnamefont
  {Kim}}, \bibinfo {author} {\bibfnamefont {D.}~\bibnamefont {Casa}}, \bibinfo
  {author} {\bibfnamefont {A.}~\bibnamefont {Said}}, \bibinfo {author}
  {\bibfnamefont {R.}~\bibnamefont {Krakora}}, \bibinfo {author} {\bibfnamefont
  {B.~J.}\ \bibnamefont {Kim}}, \bibinfo {author} {\bibfnamefont
  {E.}~\bibnamefont {Kasman}}, \bibinfo {author} {\bibfnamefont
  {X.}~\bibnamefont {Huang}}, \ and\ \bibinfo {author} {\bibfnamefont
  {T.}~\bibnamefont {Gog}},\ }\href {\doibase 10.1038/s41598-018-20396-z}
  {\bibfield  {journal} {\bibinfo  {journal} {Sci. Rep.}\ }\textbf {\bibinfo
  {volume} {8}},\ \bibinfo {pages} {1958} (\bibinfo {year} {2018})}\BibitemShut
  {NoStop}%
\bibitem [{\citenamefont {Winter}\ \emph {et~al.}(2017)\citenamefont {Winter},
  \citenamefont {Tsirlin}, \citenamefont {Daghofer}, \citenamefont {van~den
  Brink}, \citenamefont {Singh}, \citenamefont {Gegenwart},\ and\ \citenamefont
  {Valent\'{\i}}}]{Wint17}%
  \BibitemOpen
  \bibfield  {author} {\bibinfo {author} {\bibfnamefont {S.~M.}\ \bibnamefont
  {Winter}}, \bibinfo {author} {\bibfnamefont {A.~A.}\ \bibnamefont {Tsirlin}},
  \bibinfo {author} {\bibfnamefont {M.}~\bibnamefont {Daghofer}}, \bibinfo
  {author} {\bibfnamefont {J.}~\bibnamefont {van~den Brink}}, \bibinfo {author}
  {\bibfnamefont {Y.}~\bibnamefont {Singh}}, \bibinfo {author} {\bibfnamefont
  {P.}~\bibnamefont {Gegenwart}}, \ and\ \bibinfo {author} {\bibfnamefont
  {R.}~\bibnamefont {Valent\'{\i}}},\ }\href
  {http://stacks.iop.org/0953-8984/29/i=49/a=493002} {\bibfield  {journal}
  {\bibinfo  {journal} {J. Phys.: Condens. Matter}\ }\textbf {\bibinfo {volume}
  {29}},\ \bibinfo {pages} {493002} (\bibinfo {year} {2017})}\BibitemShut
  {NoStop}%
\bibitem [{\citenamefont {Moretti~Sala}\ \emph {et~al.}(2018)\citenamefont
  {Moretti~Sala}, \citenamefont {Martel}, \citenamefont {Henriquet},
  \citenamefont {Al~Zein}, \citenamefont {Simonelli}, \citenamefont {Sahle},
  \citenamefont {Gonzalez}, \citenamefont {Lagier}, \citenamefont {Ponchut},
  \citenamefont {Huotari}, \citenamefont {Verbeni}, \citenamefont {Krisch},\
  and\ \citenamefont {Monaco}}]{MorettiSala2018}%
  \BibitemOpen
  \bibfield  {author} {\bibinfo {author} {\bibfnamefont {M.}~\bibnamefont
  {Moretti~Sala}}, \bibinfo {author} {\bibfnamefont {K.}~\bibnamefont
  {Martel}}, \bibinfo {author} {\bibfnamefont {C.}~\bibnamefont {Henriquet}},
  \bibinfo {author} {\bibfnamefont {A.}~\bibnamefont {Al~Zein}}, \bibinfo
  {author} {\bibfnamefont {L.}~\bibnamefont {Simonelli}}, \bibinfo {author}
  {\bibfnamefont {C.~J.}\ \bibnamefont {Sahle}}, \bibinfo {author}
  {\bibfnamefont {H.}~\bibnamefont {Gonzalez}}, \bibinfo {author}
  {\bibfnamefont {M.-C.}\ \bibnamefont {Lagier}}, \bibinfo {author}
  {\bibfnamefont {C.}~\bibnamefont {Ponchut}}, \bibinfo {author} {\bibfnamefont
  {S.}~\bibnamefont {Huotari}}, \bibinfo {author} {\bibfnamefont
  {R.}~\bibnamefont {Verbeni}}, \bibinfo {author} {\bibfnamefont
  {M.}~\bibnamefont {Krisch}}, \ and\ \bibinfo {author} {\bibfnamefont
  {G.}~\bibnamefont {Monaco}},\ }\href {\doibase 10.1107/S1600577518001200}
  {\bibfield  {journal} {\bibinfo  {journal} {Journal of Synchrotron
  Radiation}\ }\textbf {\bibinfo {volume} {25}},\ \bibinfo {pages} {580}
  (\bibinfo {year} {2018})}\BibitemShut {NoStop}%
\bibitem [{\citenamefont {Sung}\ \emph {et~al.}(2016)\citenamefont {Sung},
  \citenamefont {Gretarsson}, \citenamefont {Proepper}, \citenamefont {Porras},
  \citenamefont {Tacon}, \citenamefont {Boris}, \citenamefont {Keimer},\ and\
  \citenamefont {Kim}}]{Sung}%
  \BibitemOpen
  \bibfield  {author} {\bibinfo {author} {\bibfnamefont {N.~H.}\ \bibnamefont
  {Sung}}, \bibinfo {author} {\bibfnamefont {H.}~\bibnamefont {Gretarsson}},
  \bibinfo {author} {\bibfnamefont {D.}~\bibnamefont {Proepper}}, \bibinfo
  {author} {\bibfnamefont {J.}~\bibnamefont {Porras}}, \bibinfo {author}
  {\bibfnamefont {M.~L.}\ \bibnamefont {Tacon}}, \bibinfo {author}
  {\bibfnamefont {A.~V.}\ \bibnamefont {Boris}}, \bibinfo {author}
  {\bibfnamefont {B.}~\bibnamefont {Keimer}}, \ and\ \bibinfo {author}
  {\bibfnamefont {B.~J.}\ \bibnamefont {Kim}},\ }\href {\doibase
  10.1080/14786435.2015.1134835} {\bibfield  {journal} {\bibinfo  {journal}
  {Philo. Mag.}\ }\textbf {\bibinfo {volume} {96}},\ \bibinfo {pages} {413}
  (\bibinfo {year} {2016})}\BibitemShut {NoStop}%
\end{thebibliography}%

\clearpage
\onecolumngrid
\begin{center}
\textbf{\large Supplemental Materials for ''Pseudospin-lattice coupling in the spin-orbit Mott insulator \sio''}
\end{center}
\setcounter{section}{0}
\setcounter{figure}{0}
\renewcommand{\thefigure}{S\arabic{figure}}
\renewcommand{\thetable}{S\arabic{table}}

\section{Sample characterization}
\sio\  single crystals were synthesized using a flux method as described in previous reports~\cite{Sung}.The sample purity and stoichiometry are of particular relevance for our magnetization measurements, as the low magnetic field region is dominated by domain wall motions that can easily cloud the effects of in-plane anisotropy (see Fig.~3(c) in Ref.~\cite{Sung}). 

X-ray Laue diffraction was used to distinguish the crystallographic directions $\mathbf{x}\equiv[1\,1\,0]$ along the Ir-O-Ir bond and $\mathbf{a} \equiv[1\,0\,0]$ along the diagonal using the conventional unit cell with $a=b=5.49\ \text{\AA}$ and $c=25.80\ \text{\AA}$. 
The magnetization measurements were performed using a commercial superconducting quantum interference device magnetometer (MPMS SQUID VSM, Quantum Design).

\section{Analysis of RIXS spectra}
\begin{figure}[h]
\includegraphics[width=0.7\linewidth]{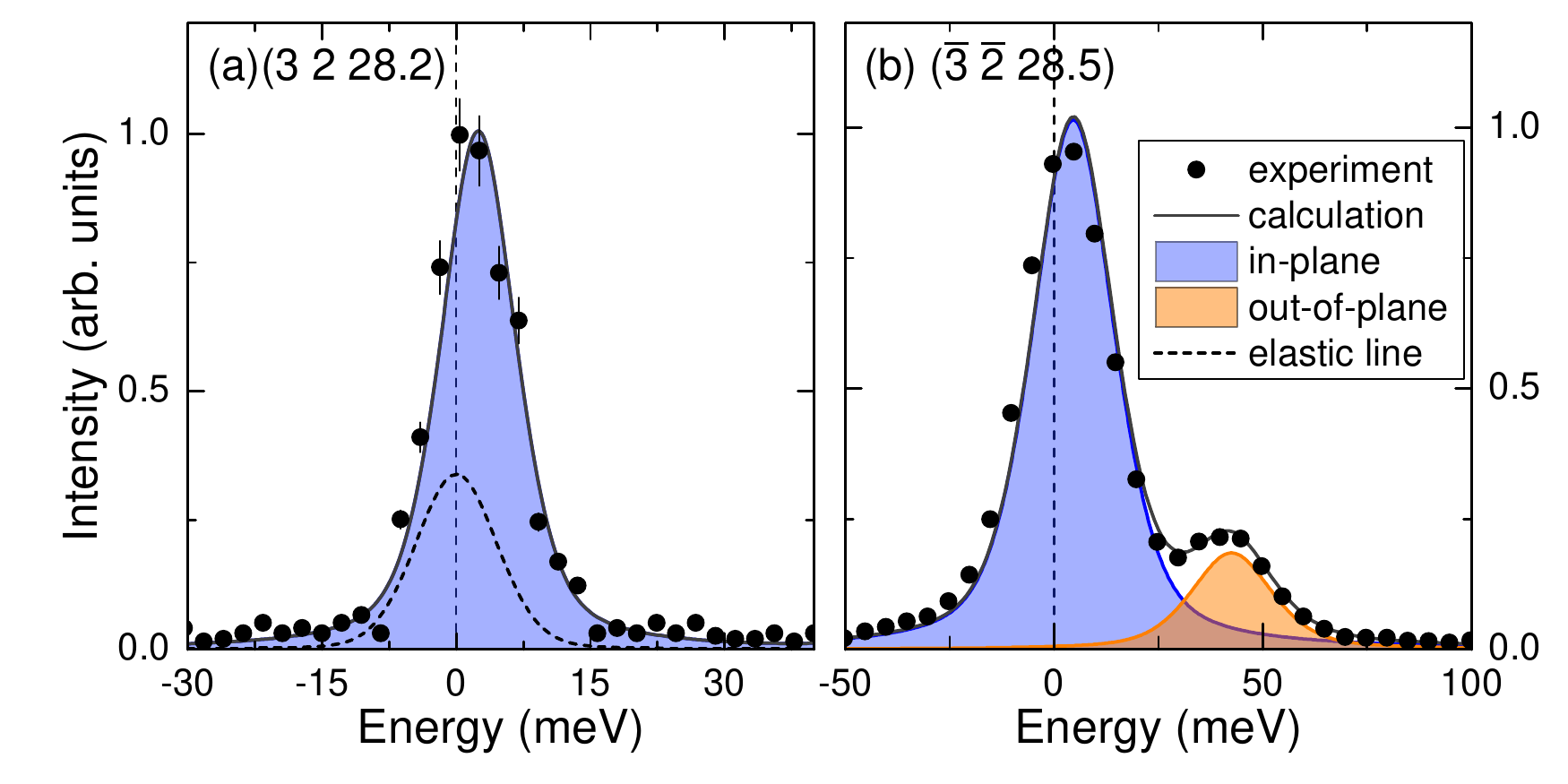}
\caption{\label{fig:SM1}Comparison of RIXS spectra taken at (a) $Q=(3\,2\,28.2)$ with $\delta E=10$ meV energy resolution (data reproduce from Fig.~\ref{fig:RIXS} in the main paper) and (b) $Q=(\bar{3}\,\bar{2}\,28.5)$ with $\delta E=23$ meV (black data points) with the calculated intensity (black line) from the model convoluted with the momentum and energy resolution of the experiments. Polarizations corrections have been applied to the calculation to match the experimental set-up. Shaded blue and orange areas indicate the nature of the modes}
\end{figure}
Figure~\ref{fig:SM1} shows resonant inelastic x-ray scattering (RIXS) data for (a) the in-plane magnetic excitation in \sio(reproduced from Fig.12 in the main text) measured at $Q=(3\,2\,28.2)$ and (b) in-plane and out-of-plane magnetic excitations measured at ID20 in ESRF~\cite{MorettiSala2018} at the almost equivalent position $Q=(\bar{3}\,\bar{2}\,28.5)$. Due to the incident polarization directions differing in the crystal frame (see next section), the two measurements have different sensitivity to in-plane and out-plane excitations. The data has good agreement with the model calculation after convolution with the experimental resolution, and reveals the presence of two distinct gaps. 

\subsection{Resonant x-ray Scattering Polarization Analysis}
Magnetic scattering in resonant x-ray scattering experiments depends on the cross-product of the incident and scattered x-ray polarization $\vec{p}_i\times\vec{p}_s$ projected to the magnetic axis. Given that the x-rays are generally produced with linear polarization in the horizontal direction of the laboratory frame, the geometry of the spectrometer and the scattering plane can be chosen to optimize the sensitivity of the experiment to different magnetic components. Additionally, a polarization analyser can be used to sort components according to their outgoing polarization.
For each experiment, the sensitivity for magnetic components along different directions is tabulated in Table \ref{tab:polarization}. 

The experiment at the Sector 6-ID-B beamline at APS was performed in vertical scattering geometry, this translates to $\sigma$ polarized photons. No polarization analyzer was used, but given that the signal is magnetic, only \spp  needs to be considered. With the restriction that the magnetic field is applied perpendicular to the scattering plane, two configurations differing by $45^\circ$ in the azimuthal angle $\psi$ with $(H\,0\,L)$ and $(H\,\bar{H}\,L)$ as the scattering planes were used
in order to apply the field along [0\,1\,0]  and [1\,1\,0] respectively. In the second configuration the measured reflections are achieved by tilting the sample around the polar angle $\chi$. 

At the P09 beamline in DESY, we used a horizontal scattering geometry giving $\pi$ polarized xrays incident at the sample. Either $\sigma'$ or $\pi'$ polarized scattered x-rays are selected by using a polarization analyzer. In order to apply the field along [1\,1\,0], the sample was mounted in the $(H\,\bar{H}\,L)$ scattering plane with a $\psi=6^\circ$ azimuthal offset to be able to reach the (4\,$\bar{5}$\,27) magnetic reflection.

For the RIXS measurement of magnetic excitations we used a horizontal scattering geometry without a polarization analyser, therefore both \psp and \ppp are present. At the sector 27-ID beamline in APS, we probed mostly in-plane polarized magnetic excitations [Fig.~\ref{fig:SM1}(a)] by adjusting $Q=(3\,2\,28.2)$ so that the incident beam is grazing the surface, as $\pi$ is nearly parallel to [0\,0\,1]. Instead the spectrum taken at ID20 in ESRF [Fig.~\ref{fig:SM1}(b)], shows both in-plane an out-of-plane polarized magnetic excitations when measuring at  $Q=(\bar{3}\,\bar{2}\,28.5)$ with nearly normal incidence.

 \begin{table*}
 \caption{\label{tab:polarization}Summary of sensitivity of resonant scattering experiments to different magnetic components.}

\begin{tabular}{|c|c|c|c|*{5}{@{\hspace{1em}}c} @{\hspace{1em}}|}
 
 \toprule
 &&&&&&&&\\ [-1.5ex]
Experiment&Scattering Plane&$\vec{Q}$&Polarization & [1\,0\,0] & [0\,1\,0] & [1\,1\,0] & [1\,$\bar{1}$\,0] & [0\,0\,1]\\[1ex]
\hline
&&&&&&&&\\ [-1.5ex]
\multirow{6}{*}{Sector 6}&\multirow{3}{*}{$\left(H\,0\,L\right)$}&(0\,1\,24)&\multirow{6}{*}{\spp} & 0.73 & 0.01 & - & - & 0.26 \\
&&(0\,1\,25)& & 0.70 & 0.01 & - & - & 0.29 \\
&&(0\,1\,26)&& 0.68 & 0.01 & - & - & 0.31 \\[1ex]
\cline{2-3}\cline{5-9}
&&&&&&&&\\ [-1.5ex]
&\multirow{3}{*}{$\left(H\,\bar{H}\,L\right)$}&(0\,1\,24)&& 0.37 & 0.24 & 0.01 & 0.60 & 0.40 \\
&&(0\,1\,25)&& 0.36 & 0.23 & 0.01 & 0.58 & 0.42 \\
&&(0\,1\,26)&& 0.34 & 0.22 & 0.01 & 0.56 & 0.44 \\[1ex]
\hline
&&&&&&&&\\ [-1.5ex]
\multirow{4}{*}{P09}&\multirow{4}{*}{$\left(H\,5/4\bar{H}\,L\right)$}&(4\,$\bar{5}$\,26)&\multirow{2}{*}{\psp} & 0.38 & 0.59 & 0.01 & 0.95 & 0.03 \\
&&(4\,$\bar{5}$\,27)&& - & - & 0.01 & 0.95 & 0.04 \\[1ex]
\cline{3-9}
&&&&&&&&\\ [-1.5ex]
&&(4\,$\bar{5}$\,26)&\multirow{2}{*}{\ppp} & 0.48 & 0.31 & 0.74 & 0.00 & 0.00 \\
&&(4\,$\bar{5}$\,27)&& - & - & 0.74 & 0.00 & 0.00 \\[1ex]
\hline
&&&&&&&&\\ [-1.5ex]
Sector 27-ID&$\left(H\,H\,L\right)$&(3\,2\,28.2)&\psp+\ppp & 0.99 & 0.93 & - & - & 0.07\\[1ex]
\hline
&&&&&&&&\\ [-1.5ex]
ID20 &$\left(H\,2/3H\,L\right)$&$(\bar{3}\,\bar{2}\,28.5)$&\psp+\ppp & 0.43 & 0.63 & - & - & 0.94\\[1ex]
\botrule
 \end{tabular}

 \end{table*}

\section{Neutron Diffraction}
\begin{figure}
\includegraphics[width=0.7\linewidth]{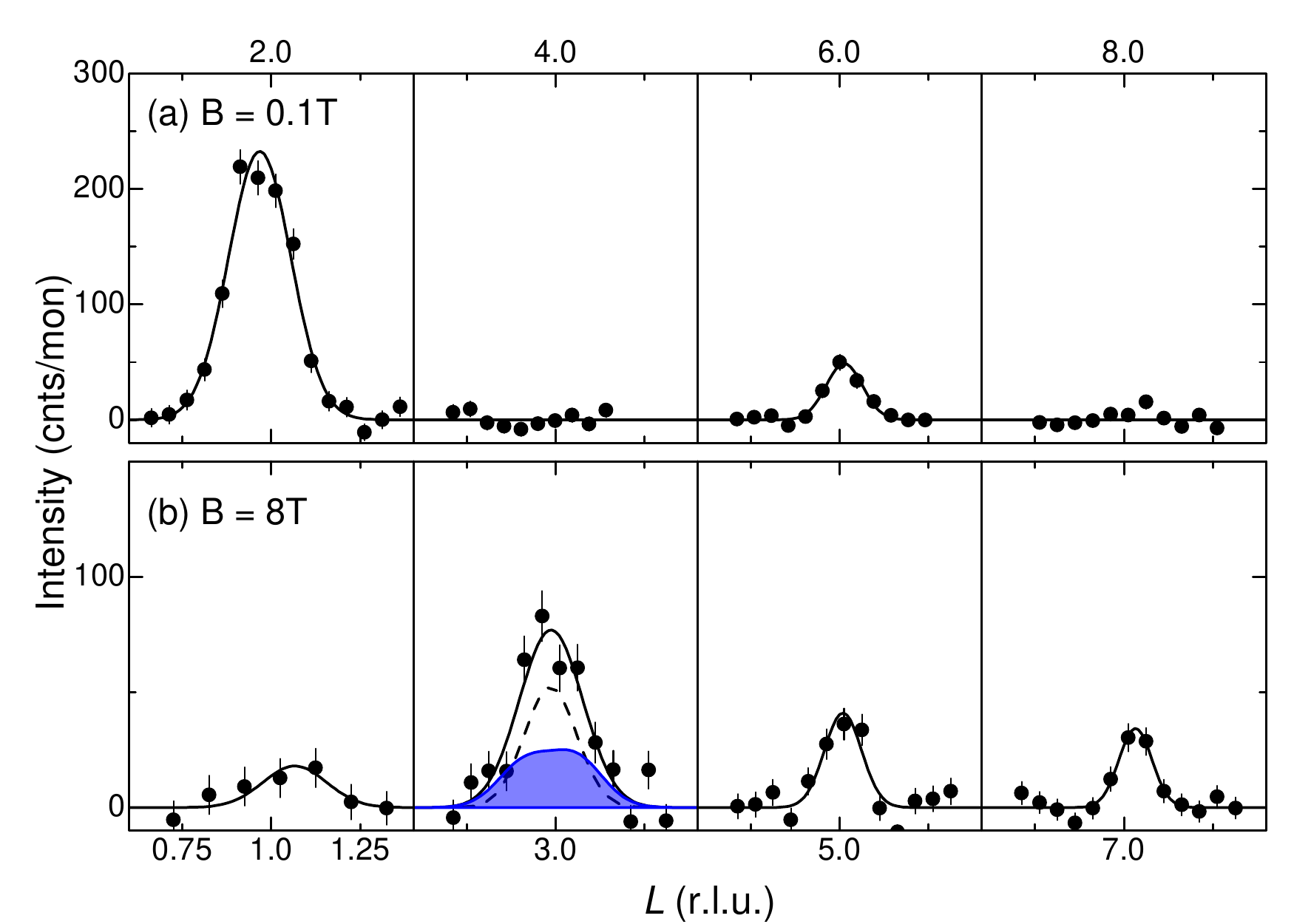}
\caption{\label{fig:SM2}Neutron elastic scattering from $(1\,0\,L)$ magnetic reflections as a function $L$ when (a)0.1T and (b)8T are applied along [0\,1\,0], solid lines are results of Gaussian fits. In the second panel in (b), the dotted line represents the contribution from the $\lambda/2$ of Al(111) reflection from the sample holder and the blue area is the contribution from the magnetic reflection.}
\end{figure}

In order to confirm the direction of the spins, stacking of the moments and domain repopulation in a magnetic field,we performed a neutron diffraction experiment in a magnetic field using the SIKA Cold Neutron 3-Axis Spectrometer of the Australian Nuclear Science and Technology Organisation, equipped with a vertical field cryomagnet and keeping $k_i=2.662\ \text{\AA}^{-1}$. In order to increase the signal and minimize neutron absorption and background, around 100 plate-like single crystals were mounted on a Si-plate. The scattering plane was set to $(H\,0\,L)$, with the magnetic field along [0\,1\,0]. The magnetic neutron cross-section is proportional to the square of the component of the magnetization perpendicular to $\vec{Q}$: for moments along [1\,0\,0], the orientation factor of $(1\,0\,L)$ increases with increasing $L$, whereas it is unaffected if the moment point along [0\,1\,0]. Fig. \ref{fig:SM2} shows the $(1\,0\,L)$ magnetic Bragg peaks with (a)0.1T and (b)8T applied magnetic field at base temperature. In (a) only $(1\,0\,4n+2)$ reflections are present, indicating \textit{uudd} stacking (with \textit{uddu} missing), the moment orientation along [0\,1\,0] can be deduced from the relative intensity of the two peaks which is affected by the magnetic form factor as well as a geometrical factor due to absorption, both decreasing the intensity with increasing $|\vec{Q}|$. Conversely in (b) the moments point along [1\,0\,0], as the \textit{uuuu} stacked $(1\,0\,2n+1)$ peaks have increasing orientation factor competing with the decreasing factors previously mentioned. Note that the $(1\,0\,3)$ peak is contaminated by an Al(111) powder-like reflection from the sample holder at $\lambda/2$ partially filtered by the PG filter. 

\end{document}